%% file: main.tex
\documentclass[%
 reprint,
 aip,
 amsmath,amssymb,
 aps,
 showkeys,
]{revtex4-2}
\input{PierrosTamplate.tex}

\usepackage[dvipsnames]{xcolor}
\usepackage{graphicx}
\usepackage{dcolumn}
\usepackage{bm}

\usepackage[colorlinks=true, urlcolor=cyan, citecolor=cyan, linkcolor=black]{hyperref}

\preprint{APS/123-QED}

\pdfoutput=1 
             
\usepackage{ifthen}
\usepackage{xspace}
\newboolean{american}
\setboolean{american}{true} 
\ifthenelse{\boolean{american}}{
\def\SorZ{z\xspace} 
\def\LLorL{l\xspace} 
}
{
\def\SorZ{s\xspace} 
\def\LLorL{ll\xspace} 
}

\usepackage{lineno}

\usepackage[T1]{fontenc} 
\usepackage{rotating} 

\usepackage{academicons}
\definecolor{orcidlogocol}{HTML}{A6CE39}

\begin{document}

\title{Functors of actions}

\author{Pierros Ntelis
}\email{pntelis -at- cppm.in2p3.fr}
\affiliation{%
 Aix Marseille Univ, CNRS/IN2P3, CPPM, Marseille, France
}%
\author{Adam Morris}%
 \email{adam.morris -at- cern.ch}
\affiliation{Universit{\"a}t Bonn - Helmholtz-Institut f{\"u}r Strahlen und Kernphysik, Bonn, Germany}

\date{\today}

\begin{abstract}
In this document, we introduce a novel formalism for any field theory and apply it to the effective field theories of large-scale structure. The new formalism is based on \textit{functors of actions} composing those theories. This new formalism predicts the \textit{actionic} fields. We discuss our findings in a \textit{cosmological gravitology} framework. We present these results with a cosmological inference approach and give guidelines on how we can choose the best candidate between those models with some latest understanding of model selection using Bayesian inference.

\end{abstract}


\maketitle

\tableofcontents

\section{Introduction}

The standard cosmological model (SMC), best described by $\Lambda$CDM parametri\SorZ{}ation, provides a satisfactory agreement with current observations~\cite{2020A&A...641A...6P}. Modified gravity (MG) is an important step in understanding models beyond the SMC~\citep{2018FrASS...5...44E,2016PhR...633....1P,2012PhR...513....1C}. Recently, effective theories  of dark energy (DE) and MG within the Hordenksi framework have been studied by \citet{PerenonMarinoniPiazza}. \citet{akrami2018neutron} have studied a doubly coupled bigravity cosmology, where the model was constrained using the detection of gravitational waves from a binary neutron star merger. These theories can been studied within a framework which we call \textit{cosmological gravitology}~\footnote{As the term suggests, it is the study of different gravitational theories within the framework of cosmology. On the other hand, we could also see the perspective in which we develop cosmology using gravitational theories. In that case we could use the terms \textit{gravitological cosmology} or \textit{gravitational cosmology}. These terms depend on what one is inspired from.}. 

At the core of these theories lies the most successful theory of gravity, general relativity (GR)~\cite{einstein1917kosmologische}. This theory assumes a four-dimensional pseudo-Riemannian manifold with a local interacting metric background that satisfies Lorentz invariance. The standard gravity action (or GR action, that contains the Einstein-Hilbert action) is given by:
\begin{equation}\label{eq:GRAction}
    S_{\rm GR} = c^4 \int d^4x \sqrt{-g} \left[ \frac{R}{16\pi G_{\rm N}} + \mathcal{L}_m \right],
\end{equation}
where $c$ is the speed of light, $g$ is the determinant of local the background metric $g_{\mu\nu}(x)$ of a massless graviton, $R(g_{\mu\nu})$ is the Ricci Scalar, $G_{\rm N}$ is the Newtonian gravitational constant and $\mathcal{L}_m$ is the Lagrangian density that describes the matter content of our universe. This Lagrangian defines the energy-momentum tensor via $T_{\mu\nu} = \frac{-2}{\sqrt{-g}} \frac{\delta S_{m}}{\delta g^{\mu\nu}}$, where $\mathcal{S}_{m}[g_{\mu\nu},...]=c^4 \int d^4x \mathcal{L}_{m}[g_{\mu\nu}(x),...]$. This field theory (FT) describes very well the large-scale structure of the universe (LSS).

In this work, we review some important MG models in a general framework and present some interesting alternative ways of thinking about the actions of effective field theories (EFTs) and, in particular, the effective field theories of large-scale structure (EFTofLSS)~\cite{carrasco2012effective}. 

As \citet{2016PhR...633....1P} reminds us, any theory of physics attempts to describe, if possible, as many observed phenomena as possible in simple mathematical laws or mathematical relationships in a unified framework. In particular, we consider field theories which do not attempt to be valid at all scales, often named EFTs. There are several ways one can apply EFT to LSS and obtain either a description of novel physics or a description of the non-linear physics.

Following \citet{2018FrASS...5...44E}, we describe the way we build the theoretical framework of an EFT of DE and MG using the equation:
\begin{equation}\label{eq:standardEFT}
	S_{\mathrm{DE,MG}} = c^4 \int d^4x \sqrt{-g} \left[ \frac{f(R)}{16\pi G_{\rm N}} + \mathcal{L}_m\left(g_{\mu\nu},\psi_m\right) \right],
\end{equation}
where  $f(R)\equiv f[R(g_{\mu\nu})]$ is the functional which has several functional forms of the Ricci scalar, namely $f(R)$ cosmologies. 
In the standard case, the matter Lagrangian matter density $\mathcal{L}_m\left(g_{\mu\nu}\right)$ contains any functional of the metric and the matter field, $\psi_m$, which includes a subset of the standard model of particle physics, namely the matter fields, and the electromagnetic interaction, at first order approximation.
The parameter space of the functional forms of $f(R)$  is large and under investigation by the community in order to effectively describe DE and gravity. Furthermore, there is substantial effort to model best the Lagrangian matter density, in order to express at best the matter of the Universe. Most theories, were investigated by experimenting with the right-hand side of~\refEq{eq:standardEFT}. In this work we experiment with the left-hand side. 

The main motivation of these theories is to explain the current physical phenomena and possibly produce explanations of current unknown issues in the standard paradigm. Additionally, these theories usually predict a new observable that has to be tested with current or future experiments so that it can be empirically confirmed. Along these lines, we propose a novel idea: we note that these theories can be reformulated, and novel formulations might lead to new understanding of the current paradigm and possibly see beyond it. The main idea discussed here is the generali\SorZ{}ation of the possible \textit{functors of actions} (FA), which generalise any FT, by introducing the set of all possible actions using functors of action. Symbolically, we define this set as $\mathcal{S}_{\rm FA}$.

Note that so far the known actions were built simply by adding several actions which have different ingredients in their integrand part, such as the following: the different Lagrangians built with different fields and symmetries; the type of the infinitesimal element of the spacetime itself; the domain of integration. In this work, we propose that we can study actions which can have different functional forms, in which there are actions which are composed by multiplications of two or more actions or even other kinds of functional form of two or more actions, such as the contraction of tensor of actions, or an integral of actions, or otherwise. As long as, we have proposed these ideas, we explore particular examples which find application in pertubation theory, and we find several models which describes new sets of equations of motions of physical systems, such as the Einstein Field Equations which describe the basic equations of motions for our Universe. This new set of Einstein Field Equations are derived from the aforementioned changes of the actions. The modifications of the actions lead also to perturbations of the actions considered for a physical system. These perturbations can be considered and interpreted as "actionic" fields.

This paper is organised as follows. In \refS{sec:Cosmological_Gravitology}, we present the philosophical framework and the main new idea in which we build these theories, in a cosmological gravitology context. We discuss the limits of these theories, using known studied actions. In \refS{sec:Constraining_an_Action_Model} we build novel actions, and in some of them we build the resulting modified EFEs, according to these novel \textit{functors of actions} theories. In \refS{sec:Physical_interpretation}, we describe interpretations of these mathematical entities. In \refS{sec:Open_Questions}, we discuss novel open questions which can be deduced with this study. Finally we conclude in \refS{sec:Conclusions}.

\section{Cosmological gravitology}\label{sec:Cosmological_Gravitology}

\begin{figure*}[ht!]
    \includegraphics[width=170mm]{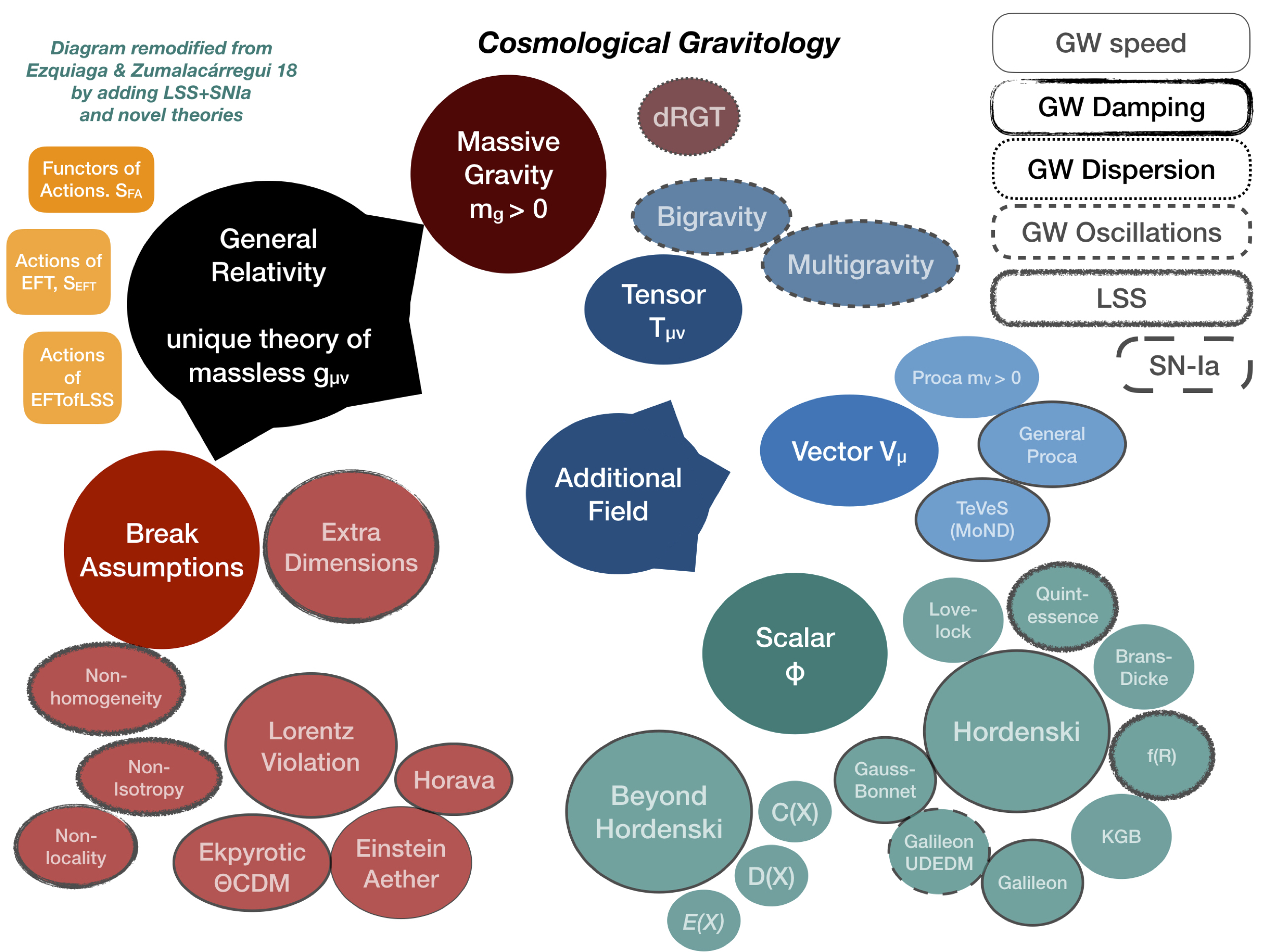} 
    \caption{\label{fig:Cosmological_Gravitology} Cosmological gravitology: in this simplified schematic, we attempt to capture the interconnections of different effective field theories and their main constraints from different observables. Diagram adapted from \citet{2018FrASS...5...44E} (see \refS{sec:Cosmological_Gravitology}). }
    \end{figure*}

Here we give a brief summary of EFTs which attempt to explain the LSS and introduce the FA framework. For a comprehensive summary of EFTs and MG in cosmology,  \citep[see][]{2018FrASS...5...44E,2016PhR...633....1P,2012PhR...513....1C}. Here, we reintroduce some of the concepts, and the interested reader can see the notation in \citet{2018FrASS...5...44E}.
We reintroduce most of these theories with an extended schematic diagram as shown in \refF{fig:Cosmological_Gravitology}. In this diagram, the different classes of theories are represented with different colors, and the subsequent classes are represented with ligher colors of the same kind. Further, we present the constraints from observational data, with different borderlines, which are represented in the right up side of this diagram.  

Standard GR is mode\LLorL{}ed with a local background metric $g_{\mu\nu}(x)$ of a massless graviton~\cite{einstein1917kosmologische}, where $x$ describes the location of the Minkowski spacetime. This theory is denoted in \refF{fig:Cosmological_Gravitology} by a black shape, and we also show how this theory is extended by several ways.

The red shapes in \refF{fig:Cosmological_Gravitology} present models that break most of the assumptions of GR, such as compactification of dimensions \cite{bailin1987kaluza,1997PhR...283..303O,1999JHEP...09..032S}, \textit{non-local} models, ekpyrotic models denoted with $\Theta$CDM, \textit{extra dimensions}~\cite{1998PhLB..436..257A,1999PhRvL..83.3370R}, \textit{Lorentz violations} \cite{2014IJMPD..2343009B}, \textit{Einstein aether} models, \textit{Horava} models~\cite{2009PhRvD..79h4008H}. The DGP models~\cite{2000PhLB..485..208D,2004JHEP...06..059N} are within this category, studying a 4D gravity in a 5D Minkowski space\cite{2000PhLB..485..208D}. Note that any model which predicts extra dimension belongs in this category, such as string theory  \cite{1999JHEP...09..032S}, Anti-de Sitter space and holography \cite{1998AdTMP...2..253W,2021arXiv210908563C} models. 

In brown, we present the \textit{massive gravity} models, $m_g > 0$, including the dRGT model, namely the resummation of massive gravity~\cite{2011PhRvL.106w1101D}.

Blue shapes denote all the models which add an additional field. These are grouped by the properties of the new field. In dark blue are models with a tensor field, $T_{\mu\nu}$. These are models such as bigravity and multigravity, which are constrained from GW oscillations. Models in light blue are the ones with a vector field $V_{\mu}$, such as Proca $m_V>0$, General Proca and TeVeS (MOND) models. In green-blue shapes, there are models with an additional scalar field, $\phi$. The most simple extensions of these models have been described by Horndeski, which we re-introduce in \refS{sec:Horndeski_Theory}. These models include Love--Lock, quintessence, Brans--Dicke, $f(R)$ \cite{1970MNRAS.150....1B,2007PhRvD..76f4004H,2004PhRvD..70d3528C}, KGB, Gauss-Bonnet, Galileon and the Galileon of unified dark energy and dark matter (UDEDM) models. The Galileon UDEDM were recently constrained by type-Ia supernovae~ \cite{2018JCAP...02..003K}, therefore it is given a green background in the diagram. Most of these models are constrained by LSS observables as indicated by the style of fuzzy grey border. The beyond-Horndeski models can by classified as the ones that have different metric within the Horndeski framework, such as $C(X)$, $D(X)$, $E(X)$, which are basically modifications of the metric. These models are not yet tested against observational data. 

\citet{carrasco2012effective} have introduced the notion of EFTofLSS. These theories have been successful on predicting some aspects of the LSS~\cite{2012JCAP...07..051B,2013JCAP...08..037P}. \citet{2013CQGra..30u4007P,2016PhRvD..93b3523B} have studied the LSS from the period of inflation to the late-time universe, using EFT. They have named these theories EFT of cosmological perturbations (EFTofCP). These theories fall under EFTofLSS since they are applied to LSS and describe physics at large scales. 

Last but not least, we introduce the newly proposed ideas to this diagram. These ideas are some basic manipulations of the action, where we introduce with a new set of any functors of actions. These manipulations may potentially produce several novel actions and potentially novel theories which include and extend current theories in which we consider only generic mathematical manipulations of the Lagrangian densities. The name we give for these theories is the \textit{functors of actions}, or FA for short. Note that this name is a basic one, and we try to give the most general name for such theories according to our understanding\footnote{Note that the name \textit{functors of actions} can be also replaced by relations of actions or any other name that best describes these novel theories. We give this name as we currently understand this best describes these mathematical entities.}. These theories are represented with orange. A possible first application is the actions of EFT or AofEFT for short, represented with light orange in this diagram. A second possible application is the one to LSS, \textit{i.e.} AofEFTofLSS, which we represent with light orange in the diagram. We describe AofEFTofLSS in \refS{sec:Functors_of_Actions}.

\subsection{Recap of Horndeski theory}\label{sec:Horndeski_Theory}
In 1974, \citet{horndeski1974second} formulated a generali\SorZ{}ed $4D$($= 3\;\mathrm{space} + 1\;\mathrm{time}$) theory of gravity. Horndeski's theory finds several applications in physics in general. From explanation of gravitational waves~\cite{babichev2017stability} to black hole models~\cite{charmousis2015lovelock,babichev2018hamiltonian,babichev2017stability}. There are several recent efforts to explain LSS using the Horndeski framework. This theory was also reformulated to contain several other paradigms, such as that of inflation, as \citet{kobayashi2011generalized} have showed.  

Modern Horndeski theories are built using the action principle:
\begin{equation}
	S_H \left[g_{\mu\nu},\phi,\psi_m \right] = \int d^4 x \sqrt{-g} \mathcal{L}_H \left[g_{\mu\nu},\phi,\psi_m \right],
\end{equation}
where $S_H$ is the Horndeski action, and $g$ is the determinant of the Jordan frame metric $g_{\mu\nu}$.

Recently, \citet{ezquiaga2017dark,kobayashi2011generalized,charmousis2015lovelock} reformulated theses class of theories, which can be consider the set of Hordenski actions, $\mathcal{S}_H$, which have up-to-second-order equations of motion. This simple Hordenski action is part of the class of Hordenski actions, and we can write symbolically $S_H \in \mathcal{S}_{H}$. This Hordenski action is defined as,
\begin{widetext}
\begin{align} \label{eq:Horndeski}
	S_H &= \int d^{4}x \sqrt{-g} \left[ \sum_{i=2}^{5} \frac{1}{8\pi G_{\rm N}}\mathcal{L}_i[g_{\mu\nu},\phi] 
+ \mathcal{L}_m\left[ g_{\mu\nu},\psi_m \right] \right],
\end{align}
\end{widetext}
with the Lagrangian densities given by:
\begin{align}
\mathcal{L}_2 &= G_2(\phi, X) \; , \\
\mathcal{L}_3 &= -G_3(\phi, X) \square\phi \; , \\
\mathcal{L}_4 &= G_4(\phi, X) R + G_{4,X}(\phi, X) \left[ (\square \phi )^2 - \phi_{;\mu\nu} \phi^{;\mu\nu} \right]  \; , \\
\mathcal{L}_5 &= G_{5}(\phi, X) G_{\mu\nu}\phi^{;\mu\nu} \nonumber \\ &- \frac{1}{6} G_{5,X}(\phi, X) \left[  (\square \phi)^3  \right. \nonumber \\ &+ \left. 2\phi_{;\mu\nu} \phi^{;\nu\alpha}\phi_{;\alpha}^{;\mu} - 3, \phi_{;\mu\nu}\phi^{;\mu\nu} \square \phi \right]  \; .
\end{align}

Here, $G_{\rm N}$ is Newton's constant, $\mathcal{L}_m$ represents the matter Lagrangian, $\psi_m$ are the matter fields, $G_{i}, i\in [2-5]$ are generic functions of a scalar field, $\phi$, and the kinetic term, $X=g^{\mu\nu}\phi_{;\mu}\phi_{;\nu}$. $R$ is the the Ricci scalar and $G_{\mu\nu}$ is the Einstein tensor. Repeated indices are summed over following Einstein's convention. Here, semicolon, ``;'',  denotes the usual covariant derivative \cite{wiki:CovariantDerivative}~$\phi_{;\mu}=\nabla_{\mu}\phi$, and comma ``,'' indicates partial derivatives $\square \phi = g^{\mu\nu} \phi_{;\mu\nu}$. The free parameters of this theory, in particular the ones from $\mathcal{L}_{4}$ and $\mathcal{L}_5$, are strongly constrained by direct measurements of the speed of GWs, (see \citet{lombriser2016breaking}). Note that $\psi_{m}$ is some simple matter field in this case.

The generic functions classify different modern Horndeski models, symbolically, as follows:
\begin{widetext}
\begin{equation}
	\mathcal{L}_H - \mathcal{L}_m \propto {\color{Orchid} G_2} {\color{red} -G_3\square\phi} {\color{Orange}+ G_4}R {\color{blue} -G_{4,X}\left\{\nabla\nabla\phi\right\}^2} {\color{OliveGreen} +G_{5}G_{\mu\nu}\phi^{;\mu\nu} - G_{5,X}\left\{ \nabla\nabla \phi \right\}^3 },
\end{equation}
\end{widetext}
where the physical interpretation of each functor is:
\begin{itemize}
	\item $\color{Orchid}G_2$: quintessence, k-essence (minimal coupling),
	\item $\color{red} G_3$: kinetic gravity braiding (derivative interactions),
	\item $\color{Orange} G_4$: generali\SorZ{}ed Brans--Dicke, $f(R)$ (non-minimal coupling),
	\item $\color{blue} G_{4,X}$: covariant Galileon (non-minimal derivative coupling),
	\item $\color{OliveGreen} G_5$: Gauss-Bonnet (non-minimal 2nd derivative coupling).
\end{itemize}

These models are constrained by current observations from LSS surveys as well as GW observables. 
There are several beyond-Horndeski schemes, among which \citet{2015JCAP...02..018G} have added some more Lagrangians to the aforementioned system, and \citet{2018FrASS...5...44E} have added modifications to the metric using the components $C(X), D(X), E(X)$.

\subsection{Functors of actions}\label{sec:Functors_of_Actions}
Here we touch on one of the fundamentals of the theoretical arguments that most field theories are built upon. 
Let us consider the set, $\mathcal{S}_{\rm FA}$, of all possible actions that can be built by any functor of actions. 
Now, instead of formulating the right-hand side of \refEq{eq:standardEFT}, we construct one basic action of field theory, namely the functors of actions, \textit{i.e.} $S_{\rm FA}$, by reformulating the left hand side of \refEq{eq:standardEFT}. Normally an action can be decomposed in a sum of a set of actions. Therefore it is easy to generalise this notion to an integral of actions according to the path integral formulation, see \citet{Peskin:1995ev}. This means that for any integral, we can promote the infinitesimal element of a real number to an infinitesimal element of real number mapped from a functional, such as the one from an action. Therefore, a new set of actions can be constructed as
\begin{equation}\label{eq:action_FA}
\boxed{
	 \mathcal{S}_{\mathrm{FA}} \ni S_{\mathrm{FA}}= \int_{\Omega_S} d S' 
	 }  \; ,
\end{equation}
where $\int_{\Omega_S}  d S'$ is an integral over all possible set of these kind of actions,   according to a path integral formalism.   In this integral we have introduced the integral domain, $\Omega_S$, and $dS'$ which is the differential of a variable action $S'$. Note that the action is a functional which takes several functions and assigns them to a real number, with units the units of an action. Therefore the actions can be used to define an integral of a set of actions.  The integral described by \refEq{eq:action_FA} can take any form, but we can assume also that this is the standard Riemann integral for the rest of this analysis. \textit{We stress here that the above expression means that the infinitesimal element $dS'$ consists of a infinitesimal action, which can be built upon any metric from any manifold.} Note that in the limit where this integral becomes a sum of series of actions then we can retrieve the standard method in which we build the action, which is the sum of individual actions, describing topology and matter, \textit{i.e.} the GR action, described by \refEq{eq:GRAction}.
What is left to define is the structure of the domain of the integral. This can take several forms. It can have an infinite set and a finite set of actions already studied in the literature, and/or an infinite set or finite set of exotic actions which are not yet studied in the literature. In a simple case, we can build this domain from several known studied action, with the following two lower and upper bounds: the upper bound element is the action of a known system, $S_K$; the lower bound element is  the action of an exotic system with a negative sign, $-S_E$; while we have a number of known and exotic actions in between the two bounds. Therefore, the integral domain structure is a simple one and we write $\Omega_S = \left[ - S_E, ..., S_K\right]$.
The integral becomes simpler with this domain, and we can write:
\begin{equation}
    S_{\mathrm{FA}}^{\rm Simple,1} = \int_{\Omega_S} d S' =  \int_{-S_E}^{S_K} d S' = S_K + S_E \; .
\end{equation} 
This means that by specifying the domain and the integrand of the integral of actions, we can rebuild the action of known and unknown new models. This method basically generali\SorZ{}es the way we construct the actions, and therefore we can build other actions than the ones we have used so far. Depending on the domain and the integrand of the integral of actions, we can build several possible \textit{functors of actions}.

Note that $\mathcal{S}_{\rm FA}$ set can contain any action which is used to build a FT or an EFT or other, \textit{i.e.} $\mathcal{S}_{\rm FA} \supset \mathcal{S}_{\rm FT} \supset \mathcal{S}_{\rm EFT}$ etc.
For example, we can imagine some simple functional forms that the action of an EFT, $S_{\mathrm{EFT}}$, can have, such that of a simple integral:
\begin{equation}\label{eq:IntegralS}
	 \mathcal{S}_{\mathrm{EFT}} \ni S_{\mathrm{EFT}} = a_S \int_{\Omega^{S}_S} d S' S' \; ,
\end{equation}
where $\Omega^{S}_S$ integral domain is a subset of $\Omega_S$, \textit{i.e.} $\Omega_S\supset \Omega_{S}^{S}$, describing all the actions resulting from \refEq{eq:IntegralS} and $a_S$ is a proportionality constant which makes the resulting object to have units of the action. \refEq{eq:IntegralS} describes the actions of EFT, which is a subset of the FA theories set, $\mathcal{S}_{\mathrm{FA}}$, see \refEq{eq:action_FA}. 
Now, we play a bit further, and we assume that there are different functionals of these actions which are written as:
\begin{equation}\label{eq:action_EFT_extended}
	\mathcal{S}_{\mathrm{FA}} \ni  \int_{\Omega_S^{F+L}} d S' \left\{  F[S'] + L[S']  \right\}  \; .
\end{equation}
where $F[S']$ and $L[S']$ are some generic functionals of the elements  $S'$ of the action set defined by the integral domain, $\Omega_S^{F+L}$, another subset of $\Omega_{S}$, \textit{i.e.} $\Omega_S\supset \Omega_{S}^{F+L}$. Some of these terms might also include terms of the form of:
\begin{equation}
	\mathcal{S}_{\mathrm{FA}} \ni \int_{\Omega_S} d S' S'_{\mu_1\, ...\, \mu_r} S'^{\mu_1\, ...\, \mu_r} \; ,
\end{equation}
where $S'_{\mu_1\, ...\, \mu_r}$ can be as simple as a tensor  with rank $r$ of some action element $S'$, or as complicated as some kind of a Riemannian tensor of some action element $S'$, which has some topological structure.

Now we simplify one of the actions as expressed by \refEq{eq:action_EFT_extended} as:
\begin{equation}\label{eq:action_EFT_forMCMC}
	S_{\mathrm{FA}}^{\mathrm{Simple, 2}} =  \int_{\Omega_S} d S'  \left[  \alpha^{(S')} S' + \beta^{(S')}  + \gamma^{(S')} ( S' )^3  \right] \; .
\end{equation}
where $\alpha^{(S')}$, $\beta^{(S')}$, $\gamma^{(S')}$ is a parametri\SorZ{}ation of one of the functionals, \textit{i.e.} the functional $F$, each one attached to the corresponding element $S'$. We omit the functional $L$ for simplification.

In the sections \ref{sec:GR_limit} and \ref{Current_universe_limit}) we discuss the limits of these theories, \textit{i.e.} the theories governed by \refEq{eq:action_FA}, using known studied actions. In \refS{sec:Constraining_an_Action_Model} we build novel actions, and in some of them we build the resulting modified EFEs, according to these novel \textit{functors of actions} theories.

\begin{widetext}
\subsubsection{GR limit}\label{sec:GR_limit}
Note that in the limit that the integral over the set of actions becomes a sum of a discrete number of actions, we recover the simple theory of GR. In particular, we have that 
\refEq{eq:action_FA} reduces to \refEq{eq:GRAction} if we assume that the integral of possible actions, $\int_{\Omega_S} dS' $, reduces to a simple sum of actions, $\sum_{i={1,2}} S_i$, and therefore we write:

\begin{align*}
\mathcal{S}_{\mathrm{FA}} \ni	\int_{\Omega_S} d S' \rightarrow \int_0^{S} d S' \stackrel{\text{GR limit}}{--\rightarrow} \int_0^{S_{\rm GR}} dS'= \sum_{i={1,2}} S_i &= S_{R} + S_{m}  \\
	  &= c^4 \int  \sqrt{-g}\frac{R}{16\pi G_{\rm N}}  d^4x
	  + c^4 \int \sqrt{-g}\,   \mathcal{L}_m\left(g_{\mu\nu}\right) d^4x  \; , \\
	 S_{\rm GR} &\equiv c^4 \int \sqrt{-g} \left[ \frac{R}{16\pi G_{\rm N}} + \mathcal{L}_m\left(g_{\mu\nu}\right) \right] d^4x \numberthis \; .
\end{align*}

\subsubsection{Limit of f(R) models}
\citet{1970MNRAS.150....1B} has introduced the f(R) models, which are simple modifications of gravity. The FA described by \refEq{eq:action_FA} has also those f(R) models as a particular limit as follows:

\begin{align*}
	\mathcal{S}_{\mathrm{FA}}  
	\supset\mathcal{S}_{\mathrm{EFT}} \ni	
	\int_0^{S} d S'  \stackrel{\text{f(R) limit}}{----\rightarrow}
	\int_0^{S_{\rm MG,1}} dS'
	= \sum_{i=1}^{2} S_i &= S_{R+f(R)} + S_m \; , \\
	S_{\rm MG,1} &\equiv \int d^{4}x \sqrt{-g} \left[ \frac{R + f(R)}{16\pi G_{\rm N}}  + \mathcal{L}_m\left[ g_{\mu\nu},\psi_m\right] \right] \numberthis \; .
\end{align*}

These models are also within the subcategory of AofEFT. Note that for realistic applications, we consider here the subset of viable low curvature f(R) models, such as the one proposed by \citet{2007PhRvD..76f4004H,2004PhRvD..70d3528C}. 

\subsubsection{Structure of integration domain of action of f(R) models}

In order to define the structure of the domain of the integration, we proceed as follows. When we would like to define the action of $f(R)$ models, we can build this model with an integral of actions, which have as a domain, $\Omega_S^{GR,f(R)}=\left\{ -S_{\rm GR}, ..., S_{f(R)} \right\}$. This means that \refEq{eq:action_FA} reduces to 
\begin{align}
     S_\mathrm{FA} \ni \int_{\Omega_S^{GR,f(R)}} dS' 
          &= 
          \int_{-S_{\rm GR}}^{S_{f(R)}} dS' \\
          &=
          \left[ S\right]_{-S_{\rm GR}}^{S_{f(R)}} \\
          &=
          S_{\rm GR}+S_{f(R)} \\
          &= \int  \sqrt{-g}  \left[ \frac{R}{16\pi G_{\rm N}} + \mathcal{L}_m \right] d^{4}x
          + 
          \int  \sqrt{-g} \frac{f(R)}{16\pi G_{\rm N}} d^4 x \\
          &=
         \int  \sqrt{-g}  \left[ \frac{R+f(R)}{16\pi G_{\rm N}} + \mathcal{L}_m \right] d^{4}x \\
         &=
         S_{R+f(R)+\mathcal{L}_m} \equiv S_{\rm MG,1}
\end{align}
Therefore, specifying the domain and the integrand of the integral of actions we can rebuild the action of $f(R)$ models. This method basically generali\SorZ{}es the way we construct the actions, and therefore we can build other actions than the ones we have used so far.

\subsubsection{Inflation limit}\label{sec:Inflation_limit}
In a similar limit where the integral of actions becomes a sum of actions, we can also recover the simplest inflationary paradigm which is described generally via the inflation action, $S_I$, \cite{STAROBINSKY198099,2020A&A...641A...6P}. In particular, we have that \refEq{eq:action_FA} reduces to:
\begin{align*}
\mathcal{S}_{\mathrm{FA}} 	 \ni		\int_{\Omega_S} d S' \stackrel{\text{Inflation limit}}{----\rightarrow} \int_0^{S_I} d S'&=  \sum_{i=1}^{2} S_i = S_1 + S_2 \; , \\
	S_{I}&\equiv \int d^{4}x \sqrt{-g} \frac{1}{16\pi G_{\rm N} } \left[ R + \frac{R^2}{6M^2} \right] \; . \numberthis
\end{align*}

\subsubsection{Horndeski limit}\label{sec:Horndeski_limit}
In a similar limit where the integral of actions becomes a sum of some actions, we can also recover the simple theory of Horndeski. In particular, we have that \refEq{eq:action_FA} reduces to the \refEq{eq:Horndeski}. This is shown simply as:
\begin{align*}
\mathcal{S}_{\mathrm{FA}} \ni \int_0^{S_H} d S' &= \sum_{i=2}^{6} S_i = \left( \sum_{i=2}^{5} S_i \right) + S_m  \; ,\\
	S_{H} &\equiv \int d^{4}x \sqrt{-g} \left[ \sum_{i=2}^{5} \frac{1}{8\pi G_{\rm N}}\mathcal{L}_i[g_{\mu\nu},\phi] + \mathcal{L}_m\left[ g_{\mu\nu},\psi_m\right] \right] \numberthis \; .
\end{align*}

\subsubsection{Actions of EFTofCP limits}
The FA theories can be also reduced to the action of EFTofCP. This relation can be mode\LLorL{}ed simply as,
\begin{align}
	\mathcal{S}_{\mathrm{FA}} &\ni \int_0^{S} d S' \stackrel{\text{EFTofCP limit}}{----\rightarrow} \int_0^{S_{\rm EFTofCP}} d S' \equiv S_{\rm EFTofCP} 
\end{align}
Note that the EFTofCP is built as:
\begin{align*}
	S_{\rm EFTofCP} &= S_m\left[ g_{\mu\nu},\Psi_i\right] + \int d^4 x \sqrt{-g} \left[ \frac{M_{*}^2}{2} f(t) R - \Lambda(t) - c(t) g^{00}  \right. \\
	&\left. + \frac{M^4_2(t)}{2} \left( \delta g^{00} \right)^2 - m^3_3(t) \delta K \delta g^{00} - m_4^2 (t) \left( \delta K^2  - \delta {K^{\mu}}_{\nu} \delta {K^{\nu}}_{\mu}  + \frac{\tilde{m}_4^2(t)}{2} ^{(3)}R \delta g^{00} \right) \right. \\
	&\left. - \bar{m}_4^2(t) \delta K^2 + \frac{\bar{m}_5(t)}{2} ^{(3)}R \delta K + \frac{\bar{\lambda}(t)}{2} ^{(3)}R^2 + ... + \frac{M_3^4(t)}{3!}  \left( \delta g^{00} \right)^3 - \frac{\bar{m}_2^3(t)}{2} \delta K  + ... \right], \numberthis
\end{align*}
where $f(t), \Lambda(t), c(t)$ are generic functions of time, $M_{*}$ is the bare Planck mass, $\delta g^{00} = g^{00} + 1$, $\delta K_{\mu\nu}$ is the perturbation of the extrinsic curvature of the constant time slices, $t$, $\delta K$ is its trace, and $^{(3)}R$ is the three-dimensional Ricci scalar.
See \citet{2013CQGra..30u4007P} for more details.
Note that this action is even more general than the Hordenksi action and also yields second-order equations of motions, as shown in the same reference. 

\subsubsection{Actions of EFTofLSS limits}
Here we give an example of how the FA theories contain the EFTofLSS as expressed in \citet{2012JCAP...07..051B}. The authors have shown that the EFTofLSS is basically expressed via an ``effective stress-energy via Einstein'' deductive approach. Using this approach, one defines the UV-IR coupling of cosmological fluctuations as arising from a reorgani\SorZ{}ation of the Einstein field equations (EFE). In particular, the Einstein tensor is decomposed in a homogeneous background (denoted by a bar) and terms that are linear (L) and non-linear (NL) in the metric perturbations, collectively denoted by $\delta X(t,\vec{x}) = X^{\rm L}(t,\vec{x}) - \bar{X}(t)$. The EFE are rewritten as: 
\begin{equation}
	\bar{G}_{\mu\nu}\left[\bar{X}\right] + \left(G_{\mu\nu}\right)^{\rm L}\left[ \delta X \right] + \left(G_{\mu\nu}\right)^{\rm NL}\left[ \delta X^2 \right] = \frac{8\pi G_{\rm N}}{c^4} T_{\mu\nu} \; .
\end{equation}
Note that the linear background equation, \textit{i.e.} $\bar{G}_{\mu\nu}\left[\bar{X}\right] = \frac{8\pi G_{\rm N}}{c^4} \bar{T}_{\mu\nu}$ and the lineari\SorZ{}ed EFE, $ \left(G_{\mu\nu}\right)^{\rm L}\left[ \delta X \right] = \frac{8\pi G_{\rm N}}{c^4} \left( T_{\mu\nu}\right)^{\rm L} $ are defined in the standard way. While the non-linear EFE can be written in a form which is similar to the linear EFE, \textit{i.e.}:
\begin{equation}
	\bar{G}_{\mu\nu}\left[\bar{X}\right] = \frac{8\pi G_{\rm N}}{c^4} \left( \tau_{\mu\nu} - \bar{T}_{\mu\nu} \right) \; ,
\end{equation}
where they have defined the effective stress-energy pseudo-tensor:
\begin{equation}
	\tau_{\mu\nu} \equiv T_{\mu\nu} - \frac{ c^4 \left( G_{\mu\nu}\right)^{\rm NL} }{8\pi G_{\rm N} } \; .
\end{equation}
Therefore, we can deduce the above formalism in a modification of the Einstein-Hilbert action as:

\begin{equation}\label{eq:EFTofLSS_action}
S_{\rm EFTofLSS} = c^4 \int d^4x \sqrt{-g} \left[ \frac{R}{16\pi G_{\rm N}} + f^{\rm L}(R) + f^{\rm NL}(R) + \mathcal{L}_m\left(g_{\mu\nu},...\right) \right].
\end{equation}
From the \refEq{eq:EFTofLSS_action}, one can perform a variational principle, \textit{i.e.} 
\begin{equation}
	\delta S_{\rm EFTofLSS}= 0 \Leftrightarrow \frac{\delta S_{\rm EFTofLSS} }{\delta g_{\mu\nu}} = 0 \; ,
\end{equation}
from which the following mathematical correspondence is inferred:
\begin{align} 
	\frac{1}{\sqrt{-g}}\frac{\delta}{\delta g_{\mu\nu}} \left[  \frac{\sqrt{-g}R}{16\pi G_{\rm N}} \right] &= \bar{G}_{\mu\nu}\left[\bar{X}\right] \\
		\frac{1}{\sqrt{-g}}\frac{\delta}{\delta g_{\mu\nu}} \left[\sqrt{-g}f^{\rm L}(R)\right] &= \left(G_{\mu\nu}\right)^{\rm L}\left[ \delta X \right] \\ 
		\frac{1}{\sqrt{-g}}\frac{\delta}{\delta g_{\mu\nu}} \left[\sqrt{-g}f^{\rm NL}(R)\right] &=\left(G_{\mu\nu}\right)^{\rm NL}\left[ \delta X^2 \right] \\
		\frac{1}{\sqrt{-g}}\frac{\delta \left[ \sqrt{-g}\mathcal{L}_m\right]}{\delta g_{\mu\nu}} &=- T_{\mu\nu}/2 \; .
\end{align}

Therefore, our new formalism basically contains the EFTofLSS in the following way. 
\begin{align}
\mathcal{S}_{\rm FA} \ni  \int_0^{S_{\rm EFTofLSS}} dS'  = &\sum_{i=1}^{4} S_i \\
&= S_{R} + S_{f^{\rm L}(R)} + S_{f^{\rm NL}(R)} + S_{m} .\\
 S_{\rm EFTofLSS} &\equiv c^4 \int d^4x \sqrt{-g} \left[ \frac{R}{16\pi G_{\rm N}} + f^{\rm L}(R) + f^{\rm NL}(R) + \mathcal{L}_m\left(g_{\mu\nu},...\right) \right].
\end{align}
Note that in practice the EFTofLSS has been formulated by adding additional terms to the standard power spectrum which correct the theoretical prediction according to 1-loop and 2-loop order for the non-linear physics predicted and motivated by EFTofLSS~\cite{2020JCAP...06..001C}.


\subsubsection{Strings limits}
\citet{polyakov1981quantum} has studied the action of string-theory dynamics \citep{Deser:1976eh,Brink:1976sc} and successfully quanti\SorZ{}ed string theory. Here we show that FA is also reduced to the one of the actions of string theory, simply as:
\begin{equation}
	\mathcal{S}_{\mathrm{FA}} \ni	 \int_0^{S} d S' \stackrel{\text{Strings limit}}{----\rightarrow}  \int_0^{S_{\rm string}} d S'= S_{\rm string} = \frac{T}{2}\int d^2\sigma \sqrt{-h} h^{ab} g_{\mu\nu}(x) \partial_a x^{\mu}(\sigma) \partial_bx^{\nu}(\sigma) \; ,
\end{equation}
where $T$ is the string tension, $g_{\mu\nu}$ is the metric of any targeted manifold of a D-dimensional space and $x_{\mu}(\sigma)$ is the coordinate of the targeted manifold. Moreover, $h_{ab}$ is the worldsheet metric, ($h^{ab}$ is its inverse), and $h$ is, as usual, the determinant of $h_{ab}$. The signatures of the metrics are chosen so that the timelike directions are positive while the spacelike directions are negative. The spacelike coordinate is denoted with $\sigma$, while the timelike coordinate is denoted with $\tau$. 

\subsubsection{Higgs action limit}\label{sec:HiggsActionLimit}
Note that FA is reduced also to the action of Higgs or any other matter action as follows. As~\cite{PhysRevLett.13.508,Higgs:1964ia,PhysRevLett.13.321} have shown, the action describing the Higgs field and its simple interactions with some fields is:
\begin{equation}
	S_{\rm m, Higgs} = c^4 \int d^4x \sqrt{\eta} \mathcal{L}_{\rm Higgs}\left[\eta_{\mu\nu},\vec{\phi}(\eta),A_{\mu} \right] \; ,
\end{equation}
where $\eta$ is the determinant of the Minkowski metric, $\eta_{\mu\nu}$, which is taken as, $-+++$,  $\vec{\phi}(\eta)$ is a vector of real scalar fields, $A_{\mu}$ is a real vector field used for the interactions, and $\mathcal{L}_{\rm Higgs}$ is the Higgs Lagrangian which is constructed using the aforementioned quantities. See \refApp{sec:HiggsAction} for more details. 
Therefore, using our formalism, we have that FA has another limiting case, the Higgs action. This can be expressed as:
\begin{equation}
	\mathcal{S}_{\mathrm{FA}} \ni	\int_{\Omega_S} d S' \rightarrow \int_0^{S} d S' \stackrel{\text{Higgs action limit}}{----\rightarrow} \int_0^{S_{\rm m, Higgs}} d S' \equiv S_{\rm m, Higgs} \; .
\end{equation}

\subsubsection{Current universe limit}\label{Current_universe_limit}
Note that since FA contains all possible actions, then ones that we have studied, as well as the exotic ones that we have not discovered yet, $\Omega_S^{KE}$, then the full action which describes the universe as a whole, $S_{\rm FA,U}$, which will be constucted via a space of actions of the whole universe, $\Omega_S^{U}=\left\{-\alpha_{\rm exotic}S_{\rm exotic},\, ..., \alpha_{H}S_{ H}\right\}$, assuming that the universe at very large scales is described by the healthy Hordenksi theories. The universe would be described by an action which is given by the actual action which describes the universe at the very large scales and high energies to the very small scales and low energies. Therefore it will include actions such as the Hordenski action, $S_H$, as well as an exotic action, $S_{\rm exotic}$. These will be the limit of the action describing the whole universe which will include the reduced Hordenski action as defined earlied, $S_{H(2-5)}$, the action of the total matter of the universe, $S_m$, which includes basically the actions of the dark matter particles, $S_{ cdm}$,  and the actions of individual galaxies at smaller scales, $S_g$. The total action will also include the actions of black hole systems, $S_{BH}$, actions of neutron star systems, $S_{NS}$, actions of gravitational wave sources, $S_{GWS}$, actions of leptons, quarks, bosons, gluons, namely the action of the standard model particles, $S_{smp}$, the action of the Higgs, $S_{\rm m, Higgs}$, the strings action, $S_{\rm strings}$,as well as some exotic system that we have not discovered yet. Therefore we can write,

\begin{align}
    \mathcal{S}_{FA} \ni S_{\rm KE}
    &= 
\int_{\Omega_S^{KE}} dS'
= S_{\rm FA,U}
    =  \int_{\Omega_S^U} dS' 
= 
\int_{-\alpha_{\rm  exotic}S_{\rm exotic}}^{\alpha_{H}S_{H}} dS' 
= 
\alpha_{H}S_{H} + \alpha_{\rm exotic}S_{\rm exotic} \\
S_{\rm FA,U}&=
\alpha_{H(2-5)}S_{H(2-5)} 
+ \sum_{cdm=1}^{\infty} \alpha_{cdm} S_{cdm} 
+ \sum_{g=1}^{\infty} \alpha_{g} S_g \nonumber
\\
&+ \sum_{BH=1}^{\infty} \alpha_{BH}S_{BH} 
+ \sum_{NS=1}^{\infty} \alpha_{NS}S_{NS} 
+ \sum_{GWS=1}^{\infty} \alpha_{GWS}S_{GWS}
\nonumber
\\
&+ \alpha_{smp}S_{smp} 
+ \alpha_{\rm m, Higgs}S_{\rm m,Higgs}
+ \alpha_{\rm strings}S_{\rm strings} + \alpha_{\rm exotic}S_{\rm exotic}
\label{eq:FA_U}
\end{align}
 where each coefficient, $\alpha_s$ with $s\in \left\{\rm H,\, 
 H(5-2),\, 
 cdm,\, 
 g,\,
 BH,\,
 NS,\,
 GWS,\,
 smp,\,
 Higgs,\,
 strings,\,
 exotic
 \right\}$, depends on the energy, $E$, and scale, $r$, applicable for each system, and it can be modelled as a step function in which it gives $1$ at the Energy and scale ranges of applicability and $0$ elsewhere. The energy and scale range of applicability or the whole form of these coefficients can be constrained by experiments.
 Note also that this section answers to the question on how the integral of all possible actions have as a limit the already studied actions. It is easy to show that applying the variational principle to \refEq{eq:FA_U}, leads to a set of equations which describe the universe and each subsystem, with a coefficient which shows the ranges of energy and scale of applicability.
 
\end{widetext}

\subsection{Cosmological inference}\label{sec:Cosmological_Inference}

In order to properly and systematically study these theories, one would proceed as follows:

\begin{enumerate}
	\item Compute the equations of motion using the standard variational principle approach, $\delta \mathcal{S}=0$, from these theories, and
		\begin{enumerate}
			\item express the corresponding analogues of Friedmann background equations~\cite{Friedman},
			\item express analogues of GW observables,
			\item or also express GWs on these Friedmann background equations analogues.
		\end{enumerate}
	\item To confront it with the observational data, using the above methodology, one would express:
		\begin{enumerate} 
			\item the LSS clustering statistics from angle positions of the tracers and their corresponding redshifts ($\hat{\theta},\hat{\phi},\hat{z}$). These clustering statistics include n-order correlation functions and their corresponding Fourier transform n-order power spectra and the cross-correlations of different matter tracers. Also relations that express the angular distance, motion distance, volume distance would also be necessary to be computed in the analogues of these theories,  $\propto \left\{ D_A(z),D_M(z),D_V(z),H(z) \right\}$, or analogues of MG observables such as the modifications of Poisson equation, anisotropic stress or lensing potentials $\propto \left\{ \mu(z,k),\eta(z,k)\simeq\frac{\Psi}{\Phi} ,\Sigma(z,k) \right\}$ \cite{2020A&A...641A...6P}. Telescopes that can be used here are the SDSS~\cite{2005ApJ...633..560E}, DESI~\cite{aghamousa2016desi}, Euclid\cite{2016arXiv160600180A}.
			\item SNIa luminosity distance diagrams, $\propto D_L(z)$  \cite{2018JCAP...02..003K},
			\item angular correlation functions that summari\SorZ{}e the Cosmic Microwave Background maps, from telescopes such as the Planck~\cite{2020A&A...641A...6P},
			\item standard sirens observables from LIGO/VIRGO~\cite{2018LRR....21....3A_VIRGO_LIGO}, LISA/Einstein Telescope~\cite{2018CQGra..35p3001C}.
		\end{enumerate}
\end{enumerate}

To study those things systematically, it would be necessary to build simulations of these observables for specific surveys, test these observables in those simulations, then apply these observables to data, using current state-of-the-art model selection methods.

In this cosmological inference analysis, we simplify things and proceed as follows.
We start by the basics of cosmological statistical inference~\cite{2015arXiv151204985L}, which arises from a generali\SorZ{}ed Bayes' theorem:
\begin{equation}\label{eq:BayesTheoremGeneral}
	P (\theta|d, M)  = \frac{ P (d|\theta, M) P (\theta|M) } {P(d)},
\end{equation}
where $P(\theta|d, M)$ is the posterior probability (\textit{i.e.} likelihood) of the parameters of physical parameters of interest, $\theta$, given some data $d$, and some theory $M$. $P (d|\theta, M)$ is the probability of the data, $d$, given some parameters, $\theta$ and some model $M$. $P(\theta,M)$ is the prior probability of the parameters of interest, $\theta$, of the theory $M$, and $P(d)$ is the Bayesian evidence of the data.

Note that the normali\SorZ{}ing constant, namely the Bayesian evidence, is defined as: 
\begin{widetext}
\begin{equation}
\boxed{
	P(d) = \int_{\Omega_M}\int_{\Omega_\theta} P(d|\theta,M) = \int_{\Omega_{M_1}} \dots \int_{\Omega_{M_n}} \int_{\Omega_{\vec{\theta_1}}} \dots \int_{\Omega_{\vec{\theta_n}}} p\left[ d|M_1(\vec{\theta_1}),\,  \dots,\,  M_n(\vec{\theta_n}) \right],
	}
\end{equation}
\end{widetext}
where it should be noted that we have adopted the notation $\int_{x} f(x) = \int_X f(x) dx $ which implies the usual Riemann integration. The Bayesian evidence is irrelevant for parameter inference. 
Usually, the set of parameters $\theta$ can be divided in some physically interesting quantities $\phi$ and a set of nuisance parameters $n$. The posterior obtained by \refEq{eq:BayesTheoremGeneral} is the joint posterior for $\theta = (\phi,n)$. The marginal posterior for the parameters of interest can now be written as (marginali\SorZ{}ing over the nuisance parameters):

\begin{equation}\label{eq:marginal_posterior}
	\mathcal{L}(\phi | d, M) \equiv P(\phi | d, M) \propto \int P(d|\phi,n,M) P(\phi, n | M )dn \; .
\end{equation} 

This PDF is the final inference on $\phi$ from the joint posterior likelihood. The next step, in order to apprehend and exploit this information, is to explore the posterior. 

\section{Constraining an FA model}\label{sec:Constraining_an_Action_Model}
In this work, we are primarily interested in constraining the models, and selecting between them is secondary. So if we apply the cosmological inference method, \refEq{eq:marginal_posterior}, for some interesting parameters, such as the physical parameters of the model $\mathcal{S}_{\mathrm{FA}}$, we can disentangle different models which describe the building blocks of nature for large-scale structures. 

We could also consider that a physical quantity is the actual action, $\phi=S'$, (or using any general formulation of an action \refEq{eq:action_FA} \textit{i.e.} $\phi=\mathcal{S}_{\mathrm{FA}}$) using the method described in \refEq{eq:marginal_posterior}, which is basically model-selection among the different ways of building the action (\textit{i.e.} the different FAs). In practice, this is a somewhat large computational problem and fairly abstract. In order to simplify things and direct ourselves to a more realistic approach, we consider the following.

Since we are interested to constrain an FA model but we want to effectively apply it to a field theory, we start by using \refEq{eq:action_EFT_forMCMC} simplified a bit further as:
\begin{equation}\label{eq:simplified_action2}
	\mathcal{S}_{\mathrm{FA}} \ni \mathcal{S}_{\mathrm{EFT}}^{\mathrm{Simplified,2}}  =  \beta^{(S_1)} S_1  +  \alpha^{(S_2)} + S_2 ,
\end{equation}
We make an assumption here, that $S_1$ has some simple Gaussian form:
\begin{equation}
	S_1(x;\phi_1,n_1) = e^{-0.5 ( (\phi_1 x - 0.1) /n_1)^2},
\end{equation}
and 
\begin{equation}
	S_2(x;\phi_2,n_2) = \phi_2 x + n_2,
\end{equation}
where $x$ is  an arbitrary variable (that can be consider as either time, space, or energy which are ingredients of an action) and $\phi_i$, $n_i$ are some physically observed and nuisance parameters, respectively, for the model $S_i$, where $i = 1,2$.
Now we consider that $\alpha^{S_1}$, $\beta^{S_1}$, $\alpha^{S_2}$, are the variables of the model, which we assume to be in real space $\left\{ \phi_i,n_i,\beta^{(S_1)},\alpha^{(S_2)} \in \mathbb{R}^{6} \right\}$. 
These assumptions mean that this model can be written in the form:
\begin{equation}\label{eq:action_FA_Simplified_2}
	\mathcal{S}_{\mathrm{EFT}}^{\mathrm{Simplified,2}} =  \beta^{(S_1)} e^{-0.5 ( (\phi_1 x - 0.1) /n_1)^2} + \alpha^{(S_2)}+ \phi_2 x + n_2 \; .
\end{equation}
In the following, we introduce a simpler theoretical model,  simulated data, their uncertainty (see \refS{sec:Theory_sims_error}), the likelihood form (see \refS{sec:Gaussian_simplification_of_the_likelihood}) and the numerical results from the MCMC approach that we use (see \refS{sec:Numerical_Results}) to constrain such model.

    \begin{figure*}[ht!]
    \hspace{0cm}
    \includegraphics[width=160mm]{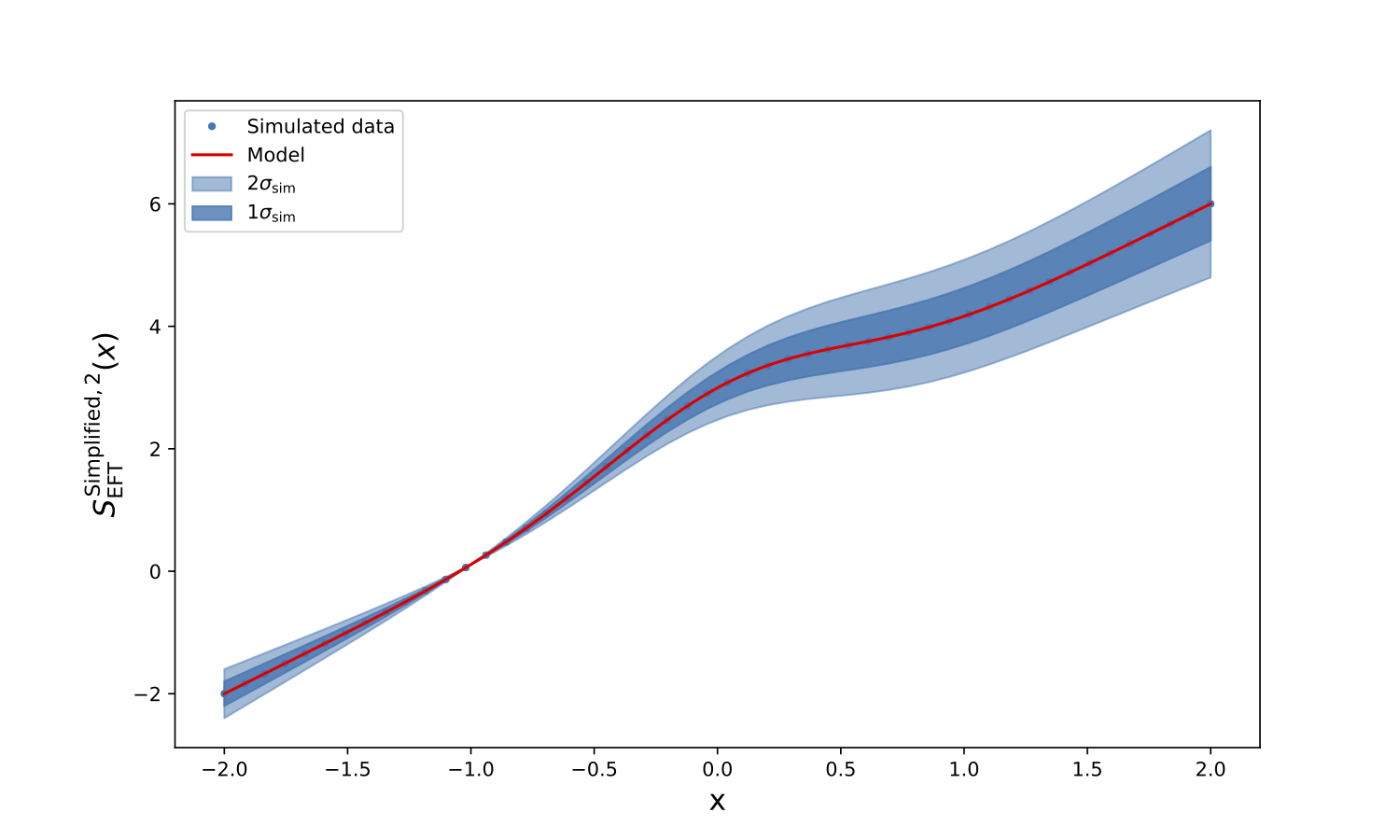} 
    \caption{\label{fig:ActionEFT_model_data_error_representation}  The simplified AofEFT model and some simulated data, as described in sections \ref{sec:Theory_sims_error} and \ref{sec:Numerical_Results}. }
    \end{figure*}

\subsection{Theoretical model, simulated data and uncertainty}\label{sec:Theory_sims_error}
We make clear that the theoretical model is described by:
\begin{equation}
	m_{\rm th}(x;\Theta^{\rm AofEFT}) = \beta^{(S_1)} e^{-0.5 ( (\phi_1 x - 0.1) /n_1)^2} + \alpha^{(S_2)} + \phi_2 x + n_2 \; ,
\end{equation}
where:
\begin{equation}\label{eq:parametrisation}
\Theta^{\rm AofEFT} = (\beta^{(S_1)},\phi_1,n_1,\alpha^{(S_2)},\phi_2,n_2) \; ,
\end{equation}
The simulated data are defined as:
\begin{equation}
	d_{\mathrm{sim}}(x;\Theta_{\rm sim}^{\rm AofEFT}) = e^{-0.5 ( (2 x - 0.1))^2} + 1 + 2 x + 0 \; ,
\end{equation}
where we have chosen that
\begin{equation}
	\Theta_{\rm sim}^{\rm AofEFT} = (\beta^{(S_1)},\phi_1,n_1,\alpha^{(S_2)},\phi_2,n_2) =(1,2,1,1,2,0) \; .
\end{equation}
The theoretical simulated uncertainty is described by a Gaussian-approximation variance, which is model as:
\begin{equation}
	\sigma_{\mathrm{sim}}(x) = m^{th}(x; 0.1 \Theta_{\rm sim}^{\rm AofEFT} ).
\end{equation}
\subsection{Gaussian simplification of the likelihood}\label{sec:Gaussian_simplification_of_the_likelihood}
We use a simple likelihood within the Gaussian-approximation limit with a diagonal covariance as:
\begin{widetext}
\begin{equation}
	-2 \ln \mathcal{L} \simeq \sum_{i}\left[ d_{\mathrm{sim}}(x_i)-m_{\mathrm{th}}(x_i;\alpha^{(S_1)},\phi_1,n_1,\beta ^{(S_2)},\phi_2,n_2) \right]^2 \sigma^{-2}_{\rm sim}(x_i) \; .
\end{equation}
\end{widetext}
Note that this is a simplification, and an interested reader can use more complex likelihoods. To sample the aforementioned likelihood, we use a modified version of \texttt{PYMC}, as it is integrated in \href{https://github.com/lontelis/cosmopit}{\texttt{COSMOPIT}}~\cite{2017JCAP...06..019N,2018JCAP...12..014N}.

\begin{table}[h!]

\caption{\label{tab:Priors} Prior information on the parametri\SorZ{}ation $\Theta^{\rm AofEFT} $. See \refS{sec:Theory_sims_error} and \refS{sec:Gaussian_simplification_of_the_likelihood}.  }

    \begin{center} 
    \begin{tabular}{l|c|c|c|c} 
    Parameter name  & [min,max] & $\mu_{\theta_i}$ & $\sigma_{\theta_i}$ & Type \\
    \hline
$\beta^{(S_1)}$  & [0.0,1.9] & - & - & Uniform \\
$\phi_{1}$  & [0.0,4] & - &- & Uniform \\
$n_1$  & [-1,1.9] & 1.3 & 0.1 & Gaussian \\
$\alpha^{(S_2)}$  & [0.1,3.0] & - & - & Uniform \\
$\phi_{2}$  & [1.1,2.9] & - &- & Uniform \\
$n_2$  & [-1,1.9] & 0.0 & 0.1 & Gaussian \\

    \hline
    \hline
    \end{tabular}
    \end{center}

\end{table}

We assume some prior information as expressed in \refT{tab:Priors} for the sampled parameters. The first column of this table shows the parameter name, the second column shows the allowed range of the parameter, the third and fourth columns shows the mean, $\mu_{\theta_i}$, and standard deviation, $\sigma_{\theta_i}$, of the Gaussian prior of this parameter, if any, and the last column shows the type of the prior parameter, which is either uniform or a Gaussian one. The reason we assume such a simple model and simple model selection is only for demonstration purposes. We leave a more realistic investigation of these kind of simulations for a future work.

\subsection{Numerical results}\label{sec:Numerical_Results}
In this section we present the numerical results of \refS{sec:Theory_sims_error} and \refS{sec:Constraining_an_Action_Model}.
In \refF{fig:ActionEFT_model_data_error_representation} we show the comparison of the model with the simulated data, assuming the choice of the values for $\Theta_{\rm sim}^{\rm AofEFT}$. This is an interesting and an initial way to constrain these kinds of models, in the case in which we do not require to compute the equations of motion by hand, and we leave another algorithm to reach the level of the FA from an observed equations of motions of a galaxy density field, or the collision of a number of elementary particles.

\begin{figure*}[ht!]
    \hspace{-1cm}
    \includegraphics[width=180mm]{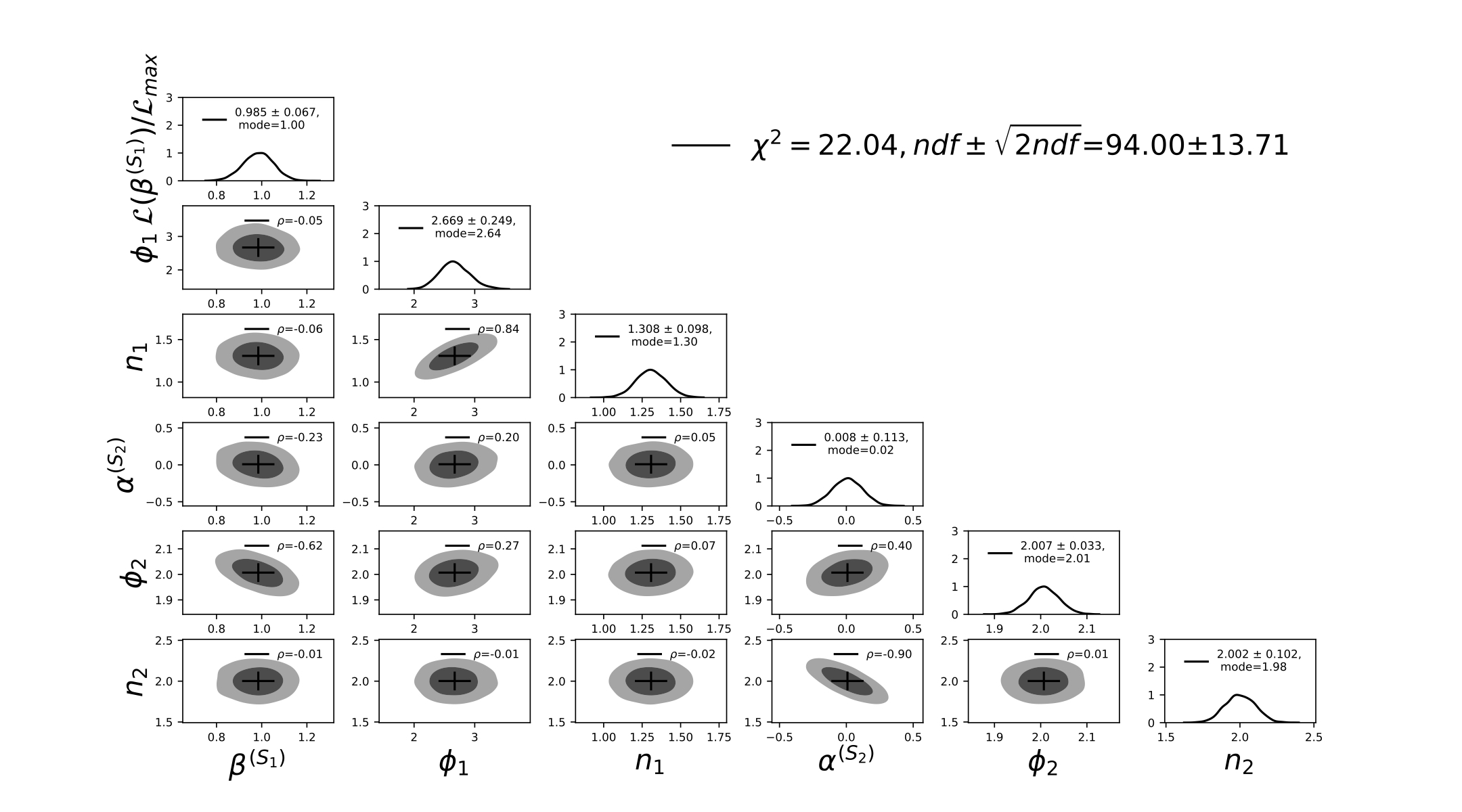} 
    \caption{\label{fig:ActionEFT_MCMCPrior_n1n2}   MCMC corner plot of the simplified AofEFT model and some simulated data. The model described here is a simple parametri\SorZ{}ation of AofEFT models, described by \refEq{eq:action_FA_Simplified_2} with the parameters described by $\Theta^{\rm AofEFT} = (\beta^{(S_1)},\phi_1,n_1,\alpha^{(S_2)},\phi_2,n_2)$. The parameters are described with JPDF of 68\% (95\%) shaded (lighted) area contours.  See sections \ref{sec:Theory_sims_error}, \ref{sec:Gaussian_simplification_of_the_likelihood} and \ref{sec:Numerical_Results}. }
    \end{figure*}

In \refF{fig:ActionEFT_MCMCPrior_n1n2}, we present the results of a Markov chain Monte Carlo (MCMC) sampling using Wilks' theorem, $\chi^2=-2\ln \mathcal{L}$~\cite{wilks}. The figure is a corner plot of the MCMC output for the parametri\SorZ{}ation of model and the likelihood $\mathcal{L}$ as described in \refS{sec:Gaussian_simplification_of_the_likelihood}. In the upper right corner-plot panel, we show the $\mathcal{L}$ as estimated from the MCMC output, as well as the corresponding number of degrees of freedom, $ndf$ and the corresponding uncertainty of the $ndf$, namely $\sqrt{2ndf}$. The diagonal of the corner matrix plot is the marginali\SorZ{}ed probability distribution of each parameter. At the top-right legend of each diagonal element of the corner matrix plot, we present the MCMC output result of each parameter with estimates of the mean and standard deviation ($\simeq 68\%$ C.L.), as well as the mode, \textit{i.e}. the value of the parameter which corresponds to the maximum value of the corresponding probability distribution function. In the off-diagonal panels of each corner plot, we present the joint probability density functions (JPDF) of the combinations of two parameters for each parametri\SorZ{}ation case. These JPDFs are described with $68\%$ (darker area) and $95\%$ (lighter area) contours. In the legend of the off-diagonal panels, we show the correlation coefficient for the combination of the two parameters, $\rho = C_{kl}/\sqrt{C_{kk} C_{ll}}$. We make our code and results, namely \texttt{AofEFT}, publicly available~\footnote{\url{https://github.com/lontelis/AofEFT}}.

\subsection{Rough constraints on FAs}
Note that these action theories, mode\LLorL{}ed by $\mathcal{S}_{\rm FA}$, are already constrained by standard observables, such as $\left\{ D_{X\in\{A,M,V,L\}}(z), H(z), \mu(z,k),\eta(z,k)\simeq\frac{\Psi}{\Phi} ,\Sigma(z,k) \right\}$ as were reviewed in \refS{sec:Cosmological_Inference}. Since these observables directly give constrains on some FA theories which are included in the reformulation of the EFTs, we therefore expect that the extended models which the FA provides are now constrained by these observables. Examples of these theories can be tracked by the EFTs, such as modifications of standard GR theories, $\mathcal{S}_{\rm FA} \supset \mathcal{S}_{\rm DE,MG}$ or the Horndeski theories, $\mathcal{S}_{\rm FA} \supset \mathcal{S}_{\rm H}$, as it is shown by \citet{2018FrASS...5...44E}.

\subsection{Relations for constraining FA models} 
As a simple example of a FA model we take the simplified AofEFT, \textit{i.e.} \refEq{eq:simplified_action2}:
\begin{equation}\tag{\ref{eq:simplified_action2}}
\boxed{
	\mathcal{S}_{\rm FA} \ni \mathcal{S}_{\mathrm{EFT}}^{\mathrm{Simplified,2}}  =  \beta^{(S_1)} S_1  +  \alpha^{(S_2)} + S_2 
	} \; .
\end{equation}
Then we can substitute the $S_1$ and $S_2$ actions with other actions that were already been discussed in the literature and define several recipes for constraining these models, with several examples given below.
Furthermore, as in most models, we use the cosmological perturbation theory, and we vary each action in study, using the variational principle, $\delta S=0$. 

\subsubsection{GR modified, 1}\label{sec:gr_modified_1}
We take for example the simplified AofEFT (\refEq{eq:simplified_action2})
and substitute $S_1$ with the first part of actions of GR theories which describe the topology, \textit{i.e.} the terms which include $R_{\mu\nu}[g_{\mu\nu}]$, while the second one will be the part of matter again from GR theories, \textit{i.e.} the terms $\mathcal{L}_{m}(g_{\mu\nu})$. Therefore we end up with:
\begin{equation}
\boxed{
	\mathcal{S}_{\mathrm{EFT}}^{\mathrm{Simplified,2, GR, 1}}  =  \beta^{(S_{R})} S_{R}  +  \alpha^{(S_m)} + S_m 
	}
	\; .
\end{equation}
Now performing the variational principle, \textit{i.e.} $\delta \mathcal{S}_{\mathrm{EFT}}^{\mathrm{Simplified,2}} = 0$, we have:
\begin{align}
	 0&= \delta S^{\rm Simplified, 2, GR, 1}   \\
	   &= \beta^{(S_{R})} \delta S_{R-2\Lambda}  + \delta a^{S_m} + \delta S_m \\
	   &= \delta \int d^4x  \sqrt{-g} \left(  \beta^{(S_{R})} \frac{c^4}{16\pi G_{\rm N}} (R-2\Lambda) +  \mathcal{L}_m \right) \\
	   &=  \int d^4x \left(   \beta^{(S_{R})} \frac{c^4}{16\pi G_{\rm N}} \delta \left[ \sqrt{-g} (R-2\Lambda) \right]  + \delta \left[ \sqrt{-g} \mathcal{L}_m \right] \right).
\end{align}
Note that as long as $\alpha^{(S_m)}$ is a constant number, the variation $ \delta \alpha^{(S_m)}$ is tautologically 0. However, we can think of a function instead of a constant number, which might result an interesting modification. We leave this for a future work. Note also that we have introduced the constant $\Lambda$ in the action of GR.
\begin{widetext}

Using simple GR definitions, we end up to the 1st modified Einstein field equation (EFE):
\begin{equation}
    \boxed{	    R_{\mu\nu}  - \frac{1}{2}Rg_{\mu\nu} + \Lambda g_{\mu\nu}  = \frac{1}{\beta^{(S_{R})}} \frac{8\pi G_{\rm N}}{c^4} T_{\mu\nu} 	}		
\end{equation}
Note that this model can be interpreted as simplified f(R)~\cite{1970MNRAS.150....1B,2007PhRvD..76f4004H,2004PhRvD..70d3528C} or nGDP~\cite{2000PhLB..485..208D,2004JHEP...06..059N} models, in which the Newtonian gravitational constant, $G_{\rm N}$, takes an effective notion and is no longer the standard constant parameter but a different one, \textit{i.e.}
\begin{equation} 
	G_{\rm eff} = \frac{1}{\beta^{(S_{R})}}G_{\rm N} \; .
\end{equation} 
Note that we have chosen to introduce the $\beta^{(S_R)}$ in the effective gravitational constant, since there are possible experiments which can falsify such theory. Possibly this rescaling can be also be introduced in the metric if there was a possibility to build a relevant experiment.
\end{widetext}

\subsubsection{GR modified, 2}\label{sec:GR_modified_2}
We also consider an additional modification of gravity as follows. We construct a simplified action, as before, but we add an extra modification of GR, by introducing an exotic action, $S_3'$. This means that this action of EFT takes the form of:
\begin{equation}
	\mathcal{S}_{\mathrm{EFT}}^{\mathrm{Simplified,2}}  =  \beta^{(S_1)} S_1  +  \alpha^{(S_2)} + S_2 + S_3^{'},
\end{equation}
and we substitute $S_1$ with the first part of actions of GR theories which describe the topology, \textit{i.e.} the terms which include $R_{\mu\nu}[g_{\mu\nu}]$, and the second one will again be the part of matter from GR theories, \textit{i.e.} the terms $\mathcal{L}_{m}(g_{\mu\nu})$. Therefore we end up with:
\begin{equation}
\boxed{
	\mathcal{S}_{\mathrm{EFT}}^{\mathrm{Simplified,2, GR, 2}}  =  \beta^{(S_{R})} S_{R}  +  \alpha^{(S_m)} + S_m + S_3^{'} 
	}\; .
\end{equation}
Performing the variational principle and applying the same procedure as before, we are left with:
\begin{widetext}
\begin{align}
	 0&= \delta S^{\rm Simplified, 2, GR, 2}_{\rm EFT}   \\
	  0 &=  \int d^4x \sqrt{-g} \delta g^{\mu\nu} \left( \beta^{(S_{R})} \frac{c^4}{16\pi G_{\rm N}} \left( R_{\mu\nu}  - \frac{1}{2}Rg_{\mu\nu} + \Lambda g_{\mu\nu} \right) -  T_{\mu\nu}/2 \right)  + \delta S_3^{'} 		\; .
\end{align}
For simplification, we assume that:
\begin{equation}
\boxed{
	\delta S_3^{'} = \int d^4x \sqrt{-g} \delta g^{\mu\nu} \delta \left[ \mathcal{L}_3 \right]_{\mu\nu} 
	}
	\; ,
\end{equation}
where $\delta \left[ \mathcal{L}_3 \right]_{\mu\nu}$ is a Lagrangian tensor which describes the new physics. These new physics can be interpreted as Lagrangian or ``actionic'' fluctuations, or Lagrangian (``actionic'') perturbations, or Lagrangian (``actionic'') waves or Lagrangian (``actionic'') fields, which are basically small fluctuations around the tensor fields describing GR and standard gravity, see \refS{sec:Physical_interpretation} for a further description.
Therefore this variational principle results to the 2nd modified EFE:
\begin{equation}
	   \boxed{ R_{\mu\nu}  - \frac{1}{2}Rg_{\mu\nu} + \Lambda g_{\mu\nu}  =\frac{1}{\beta^{(S_{R})}} \frac{8\pi G_{\rm N}}{c^4} \left( T_{\mu\nu} +  \delta \left[ \mathcal{L}_3 \right]_{\mu\nu} \right) 	} \;	.
\end{equation}
\end{widetext}
Simple DE equation of state models suggest that modification of the pressure and matter-energy result in the following relation, $w = - \rho/P $. In our case, however, this notion is generali\SorZ{}ed. In the case in which we assume that:
\begin{align}
		\delta \left[ \mathcal{L}_3 \right]_{\mu\nu} \rightarrow \delta \mathcal{L}_3
	\left(
	\begin{matrix}
	\rho & 0  & 0  & 0 \\ 
	0 & 0  & 0  & 0 \\
	0 & 0  & 0  & 0 \\
	0 & 0  & 0  & 0 \\		
	\end{matrix} \right) \;	,
\end{align}
means that the exotic Lagrangian tensor has only non-zero element the first component, which is a simplification.
Therefore to make the connection with $w$, we have that:
\begin{align}
		T^{(2)}_{\mu\nu} =
	\left( 
	\begin{matrix} 
	 \rho (1 + \delta \mathcal{L}_3)  & 0  & 0  & 0 \\ 
	0 & -P  & 0  & 0 \\
	0 & 0  & -P & 0 \\
	0 & 0  & 0  & -P
	\end{matrix} \right) \; .
\end{align}
By constraining the standard equation of state $w$, we can constrain this model of modified-GR-2 AofEFT as:
\begin{align} w=- \left(1 + \delta \mathcal{L}_3 \right)^{-1} \; ,
\end{align}
and if there is a redshift dependence in $w$, there is a redshift dependence $\delta \mathcal{L}_3$.
Therefore, this model assumes an effective gravitational Newton constant and a particular novel equation of state. For example, a measurement of a constant (non-redshift-dependent) equation of state, \textit{i.e.} $w \simeq -0.9$, corresponds to an exotic AofEFT model with $\delta \mathcal{L}_3 \simeq 0.\bar{1}$.

\subsubsection{GR modified, 3}
Taking \refEq{eq:simplified_action2}, we substitute $S_1$ with the first part of actions of Hordenksi theories which describe the topology, \textit{i.e.} the terms which include $S_{H(2-5)} \propto \int d^4 x \sqrt{-g} \sum_{i=2}^5\mathcal{L}_i$, while the second one will again be the part of matter from Horndeski theories, \textit{i.e.} the terms $S_{m} \propto \int d^4 x\mathcal{L}_{m}(g_{\mu\nu})$. Therefore we end up with:
\begin{equation}
\boxed{
	\mathcal{S}_{\mathrm{EFT}}^{\mathrm{Simplified,2}}  =  \beta^{(S_{H(2-5)})} S_{H(2-5)}  +  \alpha^{(S_m)} + S_m
	} \; .
\end{equation}
Now performing the variational principle, \textit{i.e.} $\delta \mathcal{S}_{\mathrm{EFT}}^{\mathrm{Simplified,2}} = 0$, we have:
\begin{equation}
	 \beta^{(S_{H(2-5)})} \delta S_{H(2-5)}  + \delta S_m = 0
\end{equation}
which means that now the universe is governed with the Horndeski parameters, plus the parameter which corresponds to these kind of action-like universes, \textit{i.e.} the $\beta^{(S_{H(2-5)})} $ free parameter, which will modify the Newtonian gravitational constant as:
\begin{equation}
G_{\rm eff} = \frac{1}{\beta^{(S_{H(2-5)})} } G_{\rm N} \; .
\end{equation}

\subsubsection{GR modified, 4 (quadratic)}\label{sec:Quadratic}
We build an MG model as follows. We have:
\begin{equation}
\boxed{
	\mathcal{S}_{\mathrm{EFT}}^{\mathrm{Quadratic}}  =  S_{R} +  \beta S_R^2 + S_m 
	}
	\; ,
\end{equation}
where the Einstein--Hilbert action is modified by an additional Einstein--Hilbert action in quadrature but modulated by a parameter $\beta$ so that $\beta S_R^2$ has action units. Therefore adopting now the action principle we have:
\begin{widetext}
\begin{align}
	 0&= \delta \mathcal{S}_{\mathrm{EFT}}^{\mathrm{Quadratic}}   \\
	   &= \delta S_{R} + \beta \delta \left( S_R^2 \right) + \delta S_m \\
	   &= \delta S_{R} \left( 1+ 2 \beta  S_R\right) + \delta S_m \\	   	
	   &=  \int d^4x  \left( \frac{c^4}{16\pi G_{\rm N}}   \delta\left[ \sqrt{-g}R \right] \right) 
	   \left[ 1 + 2 \beta  \int d^4x  \sqrt{-g} \left( \frac{c^4}{16\pi G_{\rm N}} R \right) \right] 
	   + \int d^4 x \delta \left[ \sqrt{-g} \mathcal{L}_m \right]  \\
	   &= \int d^4x \sqrt{-g} \delta g^{\mu\nu} \left\{ \frac{c^4}{16\pi G_{\rm N}} \left( R_{\mu\nu}  - \frac{1}{2}Rg_{\mu\nu} \right) \left[ 1 + 2 \beta  \int d^4x  \sqrt{-g} \left( \frac{c^4}{16\pi G_{\rm N}} R \right) \right] - T_{\mu\nu}/2 \right\}.
\end{align}
The expression inside the curly brackets has to be 0 in order to minimi\SorZ{}e the action, according to the action principle argument, therefore we have that the modified EFEs are given by:
\begin{equation}\label{eq:MGR_5th}
\boxed{
R_{\mu\nu}  - \frac{1}{2}Rg_{\mu\nu}  = \left[ 1 + \beta  \int d^4x  \sqrt{-g} \left( \frac{c^4}{8\pi G_{\rm N}} R \right) \right]^{-1} \frac{8\pi G_{\rm N}}{c^4} T_{\mu\nu} 
}
\; .
\end{equation}
\end{widetext}
This means that the $G_{\rm eff}$ is:
\begin{equation}
\boxed{
	G_{\rm eff} = G_{\rm N} \left[ 1 +  \beta  \int d^4x  \sqrt{-g} \left( \frac{c^4}{8\pi G_{\rm N}} R \right) \right]^{-1}  }
	\; .
\end{equation}
The term $\beta  \int d^4x  \sqrt{-g} \left( \frac{c^4}{8\pi G_{\rm N}} R \right) $ can be considered a fluctuation of the action, or in other words an ``actionic'' field, which modulates the Newtonian gravitational constant, see \refS{sec:Physical_interpretation}.
As one can see, these kinds of modifications of EFE are quite different than the ones predicted by $f(R)$, or analogously by $f(T)$, or by adding fluid terms in the Lagrangian.

For example, the $f(R)$ predicts a different modification of the EFEs, which are given by:
\begin{equation}
\frac{df(R)}{dR} R_{\mu\nu} - \frac{1}{2} f(R) g_{\mu\nu} - \left( \nabla_{\mu}\nabla_{\nu}- g_{\mu\nu}\square \right) \frac{df}{dR} = \frac{8 \pi G_N}{c^4} T_{\mu\nu} ,
\end{equation}
as \citet{2010RvMP...82..451S} have shown. 

We can also use the mini-superspace assumption, to simplify the integral. For this to work, we need to make a further assumption about the metric in consideration. 
Assuming an FLRW metric, we have that the determinant of this metric is 
$g= -c^2a^6(t)$,
where $a(t)$ is the usual scale factor, and the Ricci scalar is computed to be, $R=\frac{6}{c^2} \left[ \frac{\ddot{a}}{a} + \left(\frac{\dot{a}}{a} \right)^2\right]$. Therefore the quantity of twice the Einstein--Hilbert action can be computed as:
\begin{align}
2 S_R 
&= \frac{c^4}{8\pi G_N} \int d^4x \sqrt{-g} R \\
&= \frac{6 c^3\mathcal{V}}{8\pi G_N}  \left[ \int dt \left( a^2 \ddot{a} + \dot{a}^2a \right) \right] \label{eq:S_R_computation} ,
\end{align}
where $\mathcal{V}$ is the total volume of the spatial space under consideration. Therefore, \refEq{eq:MGR_5th} and \refEq{eq:S_R_computation}, can be used to describe a universe, with FA theory, under actionic fluctuations modulated by $\beta$.

\section{Physical interpretation: actionic fields}\label{sec:Physical_interpretation}
In sections \ref{sec:Cosmological_Gravitology} and \ref{sec:Functors_of_Actions}, we have shown how we can generali\SorZ{}e several field theories using the functors of actions approach. In \refS{sec:Constraining_an_Action_Model}, we have shown how we can expand, extend and constrain theories obtained using this method. In particular, we have shown that we can add and consider several functors that result in different equations than previous authors have considered. We have provided two concrete examples of the physical interpretation of the additional information that is generated by these novel manipulations of the actions. In particular, we have shown that we can consider the addition of a fluctuation of an action, \textit{i.e.} $\delta S$, as we described in sections \ref{sec:GR_modified_2} and \ref{sec:Quadratic}, which we call ``actionic fields'' arising from ``actionic'' fluctuations and/or perturbations.

The word ``actionic'' does not exist in the literature yet, since this is a new concept that it is introduced with this work. ``Actionic fluctuation'' or ``actionic perturbation'' is the fluctuation or perturbation of an action. ``Actionic field'' is a new compound word, which tries to capture the new concept of fields produced by perturbations or fluctuations of the action. These novel concepts can be applied to most fields of physics, which are described by equations of motions arising from an action principle, from the very small and highly energetic scales described best by the standard model of particle physics, to the largest possible scales, \textit{i.e.} cosmological scales, described best by the standard model of cosmology.
These results render the \textit{functors of actions} theories worth investigating further.

\section{Open questions}\label{sec:Open_Questions}
In this study, we introduced the generic set of all possible actions, $\mathcal{S}_{\rm FA}$, and we used simple mathematical algebra, \textit{i.e.} the use of integrals, functionals and functors to redefine the standard actions which have been studied in the literature. This opens the field of constructing actions in a more abstract way, which leads to new field theories. Along these lines, there is the question, "can we build a space of actions, $\tilde{\mathcal{S}}_{\rm FA}$, which has the set of all possible actions $\mathcal{S}_{\rm FA}$, with an added structure?" . The use of this new set of all possible actions, introduces an additional question, "can we build an algebraic structure with the aforementioned set which can be developed further?". These questions naturally lead to questions such as, "Are there any realistic actionic fields ?", "what is the best mathematical construct which describes the universe better than the ways considered before ?".  We leave the answers to these questions for future work.

\section{Conclusions}\label{sec:Conclusions} 
In this work, we briefly summari\SorZ{}e effective field theory (EFT) and cosmological perturbation theory within the cosmological gravitology framework.

We propose the novel idea of reformulating the action principle using functors of the action. This is a general way of performing variations on EFTs, and by extension it can be applied to the EFT of large-scale structures with arguments that arise from the action principle, namely functors of actions. We have shown that this method produces mathematically several alternative and complementary models to the $\Lambda$CDM model. It is also possible to extend these theories in a more concrete mathematical framework beyond what is presented in this paper. We introduce the ``actionic'' fluctuations, perturbations and fields, which are new concepts resulting from perturbations of the action quantity.
We provide guidelines for constraining these models systematically with latest cosmological inference techniques. We demonstrate how some simple classes of these models can be constrained using a Gaussian approximated likelihood analysis with some numerically simulated data and errors. We also express some relations which can be used as observables of the functors of actions. Some of these observables are related to the equation of state and the gravitational constant, therefore current constraints of these observables can constrain some of these models. 

Theories obtained using our method may be able to offer a viable alternative or be complementary to $\Lambda$CDM or modified gravity, or more generally the aforementioned effective field theories of large-scale structures. To confirm this statement a more detailed theoretical and numerical analysis is required. We plan to present such an analysis in a future work. Another possible route would be to reconsider what is beyond the variational principle, or the principle of least action, for \textit{e.g.} $\mathcal{S}\left[ \delta \mathcal{S} \right] $. 

\vspace{0.5cm}
\hspace{3.cm} 
$\mathcal{Q}. \mathcal{E}. \mathcal{D}$.


\section*{ACKNOWLEDGEMENTS}

	PN acknowledges financial support from ``Centre National d'\'Etudes Spatiales'' (CNES). 
	
	PN would like to thank F. Piazza for his inspirational phrase: ``Here, we are trying to open new possibilities!''. PN is grateful for comments on the draft from L. Amendola, J.P. Solovej, and discussion from A. Blanchard, F.H. Couannier, E.N. Saridakis, Y.Dalianis which improved the presentation of this work. 

	We acknowledge open libraries support \texttt{IPython} \cite{4160251}, \texttt{Matplotlib} \cite{Hunter:2007}, \texttt{NUMPY} \cite{Walt:2011:NAS:1957373.1957466} \texttt{SciPy 1.0} \cite{2019arXiv190710121V},  \href{https://github.com/lontelis/cosmopit}{\texttt{COSMOPIT}} \cite{2017JCAP...06..019N,2018JCAP...12..014N}.

\bibliography{apssamp}
\bibliographystyle{aipauth4-2}

\appendix

\section{The Higgs matter field}\label{sec:HiggsAction}

As \cite{PhysRevLett.13.508,Higgs:1964ia,PhysRevLett.13.321} have shown, the matter fields have one main component, the Higgs field, which gives mass to the other fields of the standard model of particle physics. In particular the action for the Higgs field is the following:
\begin{equation}
	S_{\rm m, Higgs} = c^4 \int d^4x \sqrt{\eta} \mathcal{L}_{\rm Higgs}\left[\eta_{\mu\nu},\vec{\phi}(\eta),A_{\mu} \right] 
\end{equation}
where $\eta$ is the determinant of the Minkowski metric, $\eta_{\mu\nu}$, which is taken as, $-+++$,  $\vec{\phi}(\eta)$ is a vector of real scalar fields and $A_{\mu}$ is a real vector field used for the interactions. The Lagrangian density is composed as:
\begin{widetext}
\begin{equation} 
	\mathcal{L}_{\rm Higgs}\left[\eta_{\mu\nu},\vec{\phi}(\eta),A_{\mu} \right]  = - \frac{1}{2} \left( \nabla \phi_1 \right)^2  - \frac{1}{2} \left( \nabla \phi_2 \right)^2 - V\left( \phi_1^2 + \phi_2^2 \right) - \frac{1}{4} F_{\mu\nu}F^{\mu\nu}
\end{equation}
\end{widetext}
where $\phi_i\equiv\phi_i(\eta_{\mu\nu}),\ i = 1,2$ are the two real scalar fields which interact with the $A_{\mu}$ field and $\left( \nabla \phi_i \right)^2 \equiv \nabla_{\mu}\phi_i \nabla^{\mu} \phi_i \equiv \eta_{\mu\nu} \nabla^{\mu}\phi_i \nabla^{\nu} \phi_i $ . Note that this $\nabla^{\mu}$ is different than the $\nabla^{\mu}$ in \refS{sec:Horndeski_Theory}. Here this $\nabla^{\mu}$ is defined as:
\begin{align}
	\nabla_{\mu} \phi_1 &= \partial_{\mu} \phi_1 - e A_{\mu} \phi_2 \\
	\nabla_{\mu} \phi_2 &= \partial_{\mu} \phi_2 - e A_{\mu} \phi_1 \\
	F_{\mu\nu} &= \partial_{\mu}A_{\nu} - \partial_{\nu}A_{\mu} 
\end{align}
where $e$ is a dimensionless coupling constant. Note that $\mathcal{L}_{\rm Higgs}\left[\eta_{\mu\nu},\vec{\phi}(\eta),A_{\mu} \right]$ is invariant under simultaneous gauge transformation of the first kind on $\phi_1\pm i \phi_2$ and of the second kind on $A_{\mu}$ . In the case where $V'(\phi_0^2) = 0$ and $V'' ( \phi^2_0 ) > 0$, where $\phi_0$ is the ground state of either $\phi_i$, then spontaneous breakdown of U(1) symmetry occurs. Note that we present a generic description of the Higgs field, which can be applied to more specific interactions which constitute the standard model.

\section{Statements \& declarations}
\subsection{Funding}
The authors declare that no funds, grants, or other support were received during the preparation of this manuscript.
\subsection{Competing interests}
The authors have no relevant financial or non-financial interests to disclose.
\subsection{Author contributions}
All authors contributed to the study conception and design. Material preparation, data collection and analysis were performed by Pierros Ntelis. The first draft of the manuscript was written by Pierros Ntelis and all authors commented on previous versions of the manuscript. All authors read and approved the final manuscript.
\subsection{Data availability}
The data that support the findings of this study are openly available in \href{https://github.com/lontelis/AofEFT}{\texttt{AofEFT}}.
\subsection{Publication}
This version of the article has been accepted for publication, after peer review (when applicable) but is not the Version of Record and does not reflect post-acceptance improvements, or any corrections. The Version of Record is available online at: http://dx.doi.org/10.1007/s10701-022-00628-z. Use of this Accepted Version is subject to the publisher’s Accepted Manuscript terms of use https://www.springernature.com/gp/open-research/policies/accepted- manuscript-terms.

\end{document}

%% file: PierrosTamplate.tex
\newcommand{\refEq}[1]{Eq.~\ref{#1}}
\newcommand{\refF}[1]{Fig.~\ref{#1}}

\newcommand{\refS}[1]{section~\ref{#1}}
\newcommand{\refT}[1]{table~\ref{#1}}
\newcommand{\refApp}[1]{appendix~\ref{#1}}
\newcommand{\numberthis}{\addtocounter{equation}{1}\tag{\theequation}}

   \newcommand{\aap}[0]{Astronomy and Astrophysics}

   \newcommand{\jcap}[0]{Journal of Cosmology and Astroparticle Physics}

   \newcommand{\mnras}[0]{Monthly Notices of the RAS}

   \newcommand{\physrep}[0]{Physics Reports}

%% file: main.bbl
\providecommand{\noopsort}[1]{}\providecommand{\singleletter}[1]{#1}%
\begin{thebibliography}{64}%
\makeatletter
\providecommand \@ifxundefined [1]{%
 \@ifx{#1\undefined}
}%
\providecommand \@ifnum [1]{%
 \ifnum #1\expandafter \@firstoftwo
 \else \expandafter \@secondoftwo
 \fi
}%
\providecommand \@ifx [1]{%
 \ifx #1\expandafter \@firstoftwo
 \else \expandafter \@secondoftwo
 \fi
}%
\providecommand \natexlab [1]{#1}%
\providecommand \enquote  [1]{``#1''}%
\providecommand \bibnamefont  [1]{#1}%
\providecommand \bibfnamefont [1]{#1}%
\providecommand \citenamefont [1]{#1}%
\providecommand \href@noop [0]{\@secondoftwo}%
\providecommand \href [0]{\begingroup \@sanitize@url \@href}%
\providecommand \@href[1]{\@@startlink{#1}\@@href}%
\providecommand \@@href[1]{\endgroup#1\@@endlink}%
\providecommand \@sanitize@url [0]{\catcode `\\12\catcode `\$12\catcode
  `\&12\catcode `\#12\catcode `\^12\catcode `\_12\catcode `\%12\relax}%
\providecommand \@@startlink[1]{}%
\providecommand \@@endlink[0]{}%
\providecommand \url  [0]{\begingroup\@sanitize@url \@url }%
\providecommand \@url [1]{\endgroup\@href {#1}{\urlprefix }}%
\providecommand \urlprefix  [0]{URL }%
\providecommand \Eprint [0]{\href }%
\providecommand \doibase [0]{https://doi.org/}%
\providecommand \selectlanguage [0]{\@gobble}%
\providecommand \bibinfo  [0]{\@secondoftwo}%
\providecommand \bibfield  [0]{\@secondoftwo}%
\providecommand \translation [1]{[#1]}%
\providecommand \BibitemOpen [0]{}%
\providecommand \bibitemStop [0]{}%
\providecommand \bibitemNoStop [0]{.\EOS\space}%
\providecommand \EOS [0]{\spacefactor3000\relax}%
\providecommand \BibitemShut  [1]{\csname bibitem#1\endcsname}%
\let\auto@bib@innerbib\@empty
\bibitem [{\citenamefont {{Abbott}}\ \emph {et~al.}(2018)\citenamefont
  {{Abbott}}, \citenamefont {{Abbott}}, \citenamefont {{Abbott}} \emph
  {et~al.}}]{2018LRR....21....3A_VIRGO_LIGO}%
  \BibitemOpen
  \bibfield  {author} {\bibinfo {author} {\bibnamefont {{Abbott}},
  \bibfnamefont {B.~P.}}, \bibinfo {author} {\bibnamefont {{Abbott}},
  \bibfnamefont {R.}}, \bibinfo {author} {\bibnamefont {{Abbott}},
  \bibfnamefont {T.~D.}},  \emph {et~al.},\ }\href
  {https://doi.org/10.1007/s41114-018-0012-9} {\bibfield  {journal} {\bibinfo
  {journal} {Living Reviews in Relativity}\ }\textbf {\bibinfo {volume} {21}},\
  \bibinfo {eid} {3} (\bibinfo {year} {2018})},\ \Eprint
  {https://arxiv.org/abs/1304.0670} {arXiv:1304.0670 [gr-qc]} \BibitemShut
  {NoStop}%
\bibitem [{\citenamefont {Aghamousa}\ \emph {et~al.}(2016)\citenamefont
  {Aghamousa}, \citenamefont {Aguilar}, \citenamefont {Ahlen}, \citenamefont
  {Alam}, \citenamefont {Allen}, \citenamefont {Prieto}, \citenamefont {Annis},
  \citenamefont {Bailey}, \citenamefont {Balland}, \citenamefont {Ballester}
  \emph {et~al.}}]{aghamousa2016desi}%
  \BibitemOpen
  \bibfield  {author} {\bibinfo {author} {\bibnamefont {Aghamousa},
  \bibfnamefont {A.}}, \bibinfo {author} {\bibnamefont {Aguilar}, \bibfnamefont
  {J.}}, \bibinfo {author} {\bibnamefont {Ahlen}, \bibfnamefont {S.}}, \bibinfo
  {author} {\bibnamefont {Alam}, \bibfnamefont {S.}}, \bibinfo {author}
  {\bibnamefont {Allen}, \bibfnamefont {L.~E.}}, \bibinfo {author}
  {\bibnamefont {Prieto}, \bibfnamefont {C.~A.}}, \bibinfo {author}
  {\bibnamefont {Annis}, \bibfnamefont {J.}}, \bibinfo {author} {\bibnamefont
  {Bailey}, \bibfnamefont {S.}}, \bibinfo {author} {\bibnamefont {Balland},
  \bibfnamefont {C.}}, \bibinfo {author} {\bibnamefont {Ballester},
  \bibfnamefont {O.}},  \emph {et~al.},\ }\href@noop {} {\bibfield  {journal}
  {\bibinfo  {journal} {arXiv preprint arXiv:1611.00036}\ } (\bibinfo {year}
  {2016})}\BibitemShut {NoStop}%
\bibitem [{\citenamefont {Akrami}\ \emph {et~al.}(2018)\citenamefont {Akrami},
  \citenamefont {Brax}, \citenamefont {Davis},\ and\ \citenamefont
  {Vardanyan}}]{akrami2018neutron}%
  \BibitemOpen
  \bibfield  {author} {\bibinfo {author} {\bibnamefont {Akrami}, \bibfnamefont
  {Y.}}, \bibinfo {author} {\bibnamefont {Brax}, \bibfnamefont {P.}}, \bibinfo
  {author} {\bibnamefont {Davis}, \bibfnamefont {A.-C.}}, and\ \bibinfo
  {author} {\bibnamefont {Vardanyan}, \bibfnamefont {V.}},\ }\href@noop {}
  {\bibfield  {journal} {\bibinfo  {journal} {Physical Review D}\ }\textbf
  {\bibinfo {volume} {97}},\ \bibinfo {pages} {124010} (\bibinfo {year}
  {2018})}\BibitemShut {NoStop}%
\bibitem [{\citenamefont {{Amendola}}\ \emph {et~al.}(2016)\citenamefont
  {{Amendola}}, \citenamefont {{Appleby}}, \citenamefont {{Avgoustidis}},
  \citenamefont {{Bacon}} \emph {et~al.}}]{2016arXiv160600180A}%
  \BibitemOpen
  \bibfield  {author} {\bibinfo {author} {\bibnamefont {{Amendola}},
  \bibfnamefont {L.}}, \bibinfo {author} {\bibnamefont {{Appleby}},
  \bibfnamefont {S.}}, \bibinfo {author} {\bibnamefont {{Avgoustidis}},
  \bibfnamefont {A.}}, \bibinfo {author} {\bibnamefont {{Bacon}}, \bibfnamefont
  {D.}},  \emph {et~al.},\ }\href@noop {} {\bibfield  {journal} {\bibinfo
  {journal} {ArXiv e-prints}\ } (\bibinfo {year} {2016})},\ \Eprint
  {https://arxiv.org/abs/1606.00180} {arXiv:1606.00180} \BibitemShut {NoStop}%
\bibitem [{\citenamefont {{Antoniadis}}\ \emph {et~al.}(1998)\citenamefont
  {{Antoniadis}}, \citenamefont {{Arkani-Hamed}}, \citenamefont
  {{Dimopoulos}},\ and\ \citenamefont {{Dvali}}}]{1998PhLB..436..257A}%
  \BibitemOpen
  \bibfield  {author} {\bibinfo {author} {\bibnamefont {{Antoniadis}},
  \bibfnamefont {I.}}, \bibinfo {author} {\bibnamefont {{Arkani-Hamed}},
  \bibfnamefont {N.}}, \bibinfo {author} {\bibnamefont {{Dimopoulos}},
  \bibfnamefont {S.}}, and\ \bibinfo {author} {\bibnamefont {{Dvali}},
  \bibfnamefont {G.}},\ }\href {https://doi.org/10.1016/S0370-2693(98)00860-0}
  {\bibfield  {journal} {\bibinfo  {journal} {Physics Letters B}\ }\textbf
  {\bibinfo {volume} {436}},\ \bibinfo {pages} {257} (\bibinfo {year}
  {1998})},\ \Eprint {https://arxiv.org/abs/hep-ph/9804398}
  {arXiv:hep-ph/9804398 [hep-ph]} \BibitemShut {NoStop}%
\bibitem [{\citenamefont {Babichev}\ \emph {et~al.}(2017)\citenamefont
  {Babichev}, \citenamefont {Charmousis}, \citenamefont {Esposito-Farese},\
  and\ \citenamefont {Leh{\'e}bel}}]{babichev2017stability}%
  \BibitemOpen
  \bibfield  {author} {\bibinfo {author} {\bibnamefont {Babichev},
  \bibfnamefont {E.}}, \bibinfo {author} {\bibnamefont {Charmousis},
  \bibfnamefont {C.}}, \bibinfo {author} {\bibnamefont {Esposito-Farese},
  \bibfnamefont {G.}}, and\ \bibinfo {author} {\bibnamefont {Leh{\'e}bel},
  \bibfnamefont {A.}},\ }\href@noop {} {\bibfield  {journal} {\bibinfo
  {journal} {arXiv preprint arXiv:1712.04398}\ } (\bibinfo {year}
  {2017})}\BibitemShut {NoStop}%
\bibitem [{\citenamefont {Babichev}\ \emph {et~al.}(2018)\citenamefont
  {Babichev}, \citenamefont {Charmousis}, \citenamefont {Esposito-Far{\`e}se},\
  and\ \citenamefont {Leh{\'e}bel}}]{babichev2018hamiltonian}%
  \BibitemOpen
  \bibfield  {author} {\bibinfo {author} {\bibnamefont {Babichev},
  \bibfnamefont {E.}}, \bibinfo {author} {\bibnamefont {Charmousis},
  \bibfnamefont {C.}}, \bibinfo {author} {\bibnamefont {Esposito-Far{\`e}se},
  \bibfnamefont {G.}}, and\ \bibinfo {author} {\bibnamefont {Leh{\'e}bel},
  \bibfnamefont {A.}},\ }\href@noop {} {\bibfield  {journal} {\bibinfo
  {journal} {arXiv preprint arXiv:1803.11444}\ } (\bibinfo {year}
  {2018})}\BibitemShut {NoStop}%
\bibitem [{\citenamefont {Bailin}\ and\ \citenamefont
  {Love}(1987)}]{bailin1987kaluza}%
  \BibitemOpen
  \bibfield  {author} {\bibinfo {author} {\bibnamefont {Bailin}, \bibfnamefont
  {D.}}and\ \bibinfo {author} {\bibnamefont {Love}, \bibfnamefont {A.}},\
  }\href@noop {} {\bibfield  {journal} {\bibinfo  {journal} {Reports on
  Progress in Physics}\ }\textbf {\bibinfo {volume} {50}},\ \bibinfo {pages}
  {1087} (\bibinfo {year} {1987})}\BibitemShut {NoStop}%
\bibitem [{\citenamefont {{Baumann}}\ \emph {et~al.}(2016)\citenamefont
  {{Baumann}}, \citenamefont {{Green}}, \citenamefont {{Lee}},\ and\
  \citenamefont {{Porto}}}]{2016PhRvD..93b3523B}%
  \BibitemOpen
  \bibfield  {author} {\bibinfo {author} {\bibnamefont {{Baumann}},
  \bibfnamefont {D.}}, \bibinfo {author} {\bibnamefont {{Green}}, \bibfnamefont
  {D.}}, \bibinfo {author} {\bibnamefont {{Lee}}, \bibfnamefont {H.}}, and\
  \bibinfo {author} {\bibnamefont {{Porto}}, \bibfnamefont {R.~A.}},\ }\href
  {https://doi.org/10.1103/PhysRevD.93.023523} {\bibfield  {journal} {\bibinfo
  {journal} {\prd}\ }\textbf {\bibinfo {volume} {93}},\ \bibinfo {eid} {023523}
  (\bibinfo {year} {2016})},\ \Eprint {https://arxiv.org/abs/1502.07304}
  {arXiv:1502.07304 [hep-th]} \BibitemShut {NoStop}%
\bibitem [{\citenamefont {{Baumann}}\ \emph {et~al.}(2012)\citenamefont
  {{Baumann}}, \citenamefont {{Nicolis}}, \citenamefont {{Senatore}},\ and\
  \citenamefont {{Zaldarriaga}}}]{2012JCAP...07..051B}%
  \BibitemOpen
  \bibfield  {author} {\bibinfo {author} {\bibnamefont {{Baumann}},
  \bibfnamefont {D.}}, \bibinfo {author} {\bibnamefont {{Nicolis}},
  \bibfnamefont {A.}}, \bibinfo {author} {\bibnamefont {{Senatore}},
  \bibfnamefont {L.}}, and\ \bibinfo {author} {\bibnamefont {{Zaldarriaga}},
  \bibfnamefont {M.}},\ }\href {https://doi.org/10.1088/1475-7516/2012/07/051}
  {\bibfield  {journal} {\bibinfo  {journal} {\jcap}\ }\textbf {\bibinfo
  {volume} {2012}},\ \bibinfo {eid} {051} (\bibinfo {year} {2012})},\ \Eprint
  {https://arxiv.org/abs/1004.2488} {arXiv:1004.2488 [astro-ph.CO]}
  \BibitemShut {NoStop}%
\bibitem [{\citenamefont {{Blas}}\ and\ \citenamefont
  {{Lim}}(2014)}]{2014IJMPD..2343009B}%
  \BibitemOpen
  \bibfield  {author} {\bibinfo {author} {\bibnamefont {{Blas}}, \bibfnamefont
  {D.}}and\ \bibinfo {author} {\bibnamefont {{Lim}}, \bibfnamefont {E.}},\
  }\href {https://doi.org/10.1142/S0218271814430093} {\bibfield  {journal}
  {\bibinfo  {journal} {International Journal of Modern Physics D}\ }\textbf
  {\bibinfo {volume} {23}},\ \bibinfo {eid} {1443009} (\bibinfo {year}
  {2014})},\ \Eprint {https://arxiv.org/abs/1412.4828} {arXiv:1412.4828
  [gr-qc]} \BibitemShut {NoStop}%
\bibitem [{\citenamefont {Brink}, \citenamefont {Di~Vecchia},\ and\
  \citenamefont {Howe}(1976)}]{Brink:1976sc}%
  \BibitemOpen
  \bibfield  {author} {\bibinfo {author} {\bibnamefont {Brink}, \bibfnamefont
  {L.}}, \bibinfo {author} {\bibnamefont {Di~Vecchia}, \bibfnamefont {P.}},
  and\ \bibinfo {author} {\bibnamefont {Howe}, \bibfnamefont {P.~S.}},\ }\href
  {https://doi.org/10.1016/0370-2693(76)90445-7} {\bibfield  {journal}
  {\bibinfo  {journal} {Phys. Lett. B}\ }\textbf {\bibinfo {volume} {65}},\
  \bibinfo {pages} {471} (\bibinfo {year} {1976})}\BibitemShut {NoStop}%
\bibitem [{\citenamefont {{Buchdahl}}(1970)}]{1970MNRAS.150....1B}%
  \BibitemOpen
  \bibfield  {author} {\bibinfo {author} {\bibnamefont {{Buchdahl}},
  \bibfnamefont {H.~A.}},\ }\href {https://doi.org/10.1093/mnras/150.1.1}
  {\bibfield  {journal} {\bibinfo  {journal} {\mnras}\ }\textbf {\bibinfo
  {volume} {150}},\ \bibinfo {pages} {1} (\bibinfo {year} {1970})}\BibitemShut
  {NoStop}%
\bibitem [{\citenamefont {{Caprini}}\ and\ \citenamefont
  {{Figueroa}}(2018)}]{2018CQGra..35p3001C}%
  \BibitemOpen
  \bibfield  {author} {\bibinfo {author} {\bibnamefont {{Caprini}},
  \bibfnamefont {C.}}and\ \bibinfo {author} {\bibnamefont {{Figueroa}},
  \bibfnamefont {D.~G.}},\ }\href {https://doi.org/10.1088/1361-6382/aac608}
  {\bibfield  {journal} {\bibinfo  {journal} {Classical and Quantum Gravity}\
  }\textbf {\bibinfo {volume} {35}},\ \bibinfo {eid} {163001} (\bibinfo {year}
  {2018})},\ \Eprint {https://arxiv.org/abs/1801.04268} {arXiv:1801.04268
  [astro-ph.CO]} \BibitemShut {NoStop}%
\bibitem [{\citenamefont {Carrasco}, \citenamefont {Hertzberg},\ and\
  \citenamefont {Senatore}(2012)}]{carrasco2012effective}%
  \BibitemOpen
  \bibfield  {author} {\bibinfo {author} {\bibnamefont {Carrasco},
  \bibfnamefont {J.~J.~M.}}, \bibinfo {author} {\bibnamefont {Hertzberg},
  \bibfnamefont {M.~P.}}, and\ \bibinfo {author} {\bibnamefont {Senatore},
  \bibfnamefont {L.}},\ }\href@noop {} {\bibfield  {journal} {\bibinfo
  {journal} {Journal of High Energy Physics}\ }\textbf {\bibinfo {volume}
  {2012}},\ \bibinfo {pages} {82} (\bibinfo {year} {2012})}\BibitemShut
  {NoStop}%
\bibitem [{\citenamefont {{Carroll}}\ \emph {et~al.}(2004)\citenamefont
  {{Carroll}}, \citenamefont {{Duvvuri}}, \citenamefont {{Trodden}},\ and\
  \citenamefont {{Turner}}}]{2004PhRvD..70d3528C}%
  \BibitemOpen
  \bibfield  {author} {\bibinfo {author} {\bibnamefont {{Carroll}},
  \bibfnamefont {S.~M.}}, \bibinfo {author} {\bibnamefont {{Duvvuri}},
  \bibfnamefont {V.}}, \bibinfo {author} {\bibnamefont {{Trodden}},
  \bibfnamefont {M.}}, and\ \bibinfo {author} {\bibnamefont {{Turner}},
  \bibfnamefont {M.~S.}},\ }\href {https://doi.org/10.1103/PhysRevD.70.043528}
  {\bibfield  {journal} {\bibinfo  {journal} {\prd}\ }\textbf {\bibinfo
  {volume} {70}},\ \bibinfo {eid} {043528} (\bibinfo {year} {2004})},\ \Eprint
  {https://arxiv.org/abs/astro-ph/0306438} {arXiv:astro-ph/0306438 [astro-ph]}
  \BibitemShut {NoStop}%
\bibitem [{\citenamefont {Charmousis}(2015)}]{charmousis2015lovelock}%
  \BibitemOpen
  \bibfield  {author} {\bibinfo {author} {\bibnamefont {Charmousis},
  \bibfnamefont {C.}},\ }in\ \href@noop {} {\emph {\bibinfo {booktitle}
  {{Modifications of Einstein's Theory of Gravity at Large Distances}}}}\
  (\bibinfo  {publisher} {Springer},\ \bibinfo {year} {2015})\ pp.\ \bibinfo
  {pages} {25--56}\BibitemShut {NoStop}%
\bibitem [{\citenamefont {{Chen}}, \citenamefont {{Maldacena}},\ and\
  \citenamefont {{Witten}}(2021)}]{2021arXiv210908563C}%
  \BibitemOpen
  \bibfield  {author} {\bibinfo {author} {\bibnamefont {{Chen}}, \bibfnamefont
  {Y.}}, \bibinfo {author} {\bibnamefont {{Maldacena}}, \bibfnamefont {J.}},
  and\ \bibinfo {author} {\bibnamefont {{Witten}}, \bibfnamefont {E.}},\
  }\href@noop {} {\bibfield  {journal} {\bibinfo  {journal} {arXiv e-prints}\
  ,\ \bibinfo {eid} {arXiv:2109.08563}} (\bibinfo {year} {2021})},\ \Eprint
  {https://arxiv.org/abs/2109.08563} {arXiv:2109.08563 [hep-th]} \BibitemShut
  {NoStop}%
\bibitem [{\citenamefont {{Clifton}}\ \emph {et~al.}(2012)\citenamefont
  {{Clifton}}, \citenamefont {{Ferreira}}, \citenamefont {{Padilla}},\ and\
  \citenamefont {{Skordis}}}]{2012PhR...513....1C}%
  \BibitemOpen
  \bibfield  {author} {\bibinfo {author} {\bibnamefont {{Clifton}},
  \bibfnamefont {T.}}, \bibinfo {author} {\bibnamefont {{Ferreira}},
  \bibfnamefont {P.~G.}}, \bibinfo {author} {\bibnamefont {{Padilla}},
  \bibfnamefont {A.}}, and\ \bibinfo {author} {\bibnamefont {{Skordis}},
  \bibfnamefont {C.}},\ }\href {https://doi.org/10.1016/j.physrep.2012.01.001}
  {\bibfield  {journal} {\bibinfo  {journal} {\physrep}\ }\textbf {\bibinfo
  {volume} {513}},\ \bibinfo {pages} {1} (\bibinfo {year} {2012})},\ \Eprint
  {https://arxiv.org/abs/1106.2476} {arXiv:1106.2476 [astro-ph.CO]}
  \BibitemShut {NoStop}%
\bibitem [{\citenamefont {{Colas}}\ \emph {et~al.}(2020)\citenamefont
  {{Colas}}, \citenamefont {{d'Amico}}, \citenamefont {{Senatore}},
  \citenamefont {{Zhang}},\ and\ \citenamefont
  {{Beutler}}}]{2020JCAP...06..001C}%
  \BibitemOpen
  \bibfield  {author} {\bibinfo {author} {\bibnamefont {{Colas}}, \bibfnamefont
  {T.}}, \bibinfo {author} {\bibnamefont {{d'Amico}}, \bibfnamefont {G.}},
  \bibinfo {author} {\bibnamefont {{Senatore}}, \bibfnamefont {L.}}, \bibinfo
  {author} {\bibnamefont {{Zhang}}, \bibfnamefont {P.}}, and\ \bibinfo {author}
  {\bibnamefont {{Beutler}}, \bibfnamefont {F.}},\ }\href
  {https://doi.org/10.1088/1475-7516/2020/06/001} {\bibfield  {journal}
  {\bibinfo  {journal} {\jcap}\ }\textbf {\bibinfo {volume} {2020}},\ \bibinfo
  {eid} {001} (\bibinfo {year} {2020})},\ \Eprint
  {https://arxiv.org/abs/1909.07951} {arXiv:1909.07951 [astro-ph.CO]}
  \BibitemShut {NoStop}%
\bibitem [{\citenamefont {{de Rham}}, \citenamefont {{Gabadadze}},\ and\
  \citenamefont {{Tolley}}(2011)}]{2011PhRvL.106w1101D}%
  \BibitemOpen
  \bibfield  {author} {\bibinfo {author} {\bibnamefont {{de Rham}},
  \bibfnamefont {C.}}, \bibinfo {author} {\bibnamefont {{Gabadadze}},
  \bibfnamefont {G.}}, and\ \bibinfo {author} {\bibnamefont {{Tolley}},
  \bibfnamefont {A.~J.}},\ }\href
  {https://doi.org/10.1103/PhysRevLett.106.231101} {\bibfield  {journal}
  {\bibinfo  {journal} {\prl}\ }\textbf {\bibinfo {volume} {106}},\ \bibinfo
  {eid} {231101} (\bibinfo {year} {2011})},\ \Eprint
  {https://arxiv.org/abs/1011.1232} {arXiv:1011.1232 [hep-th]} \BibitemShut
  {NoStop}%
\bibitem [{\citenamefont {Deser}\ and\ \citenamefont
  {Zumino}(1976)}]{Deser:1976eh}%
  \BibitemOpen
  \bibfield  {author} {\bibinfo {author} {\bibnamefont {Deser}, \bibfnamefont
  {S.}}and\ \bibinfo {author} {\bibnamefont {Zumino}, \bibfnamefont {B.}},\
  }\href {https://doi.org/10.1016/0370-2693(76)90089-7} {\bibfield  {journal}
  {\bibinfo  {journal} {Phys. Lett. B}\ }\textbf {\bibinfo {volume} {62}},\
  \bibinfo {pages} {335} (\bibinfo {year} {1976})}\BibitemShut {NoStop}%
\bibitem [{\citenamefont {{Dvali}}, \citenamefont {{Gabadadze}},\ and\
  \citenamefont {{Porrati}}(2000)}]{2000PhLB..485..208D}%
  \BibitemOpen
  \bibfield  {author} {\bibinfo {author} {\bibnamefont {{Dvali}}, \bibfnamefont
  {G.}}, \bibinfo {author} {\bibnamefont {{Gabadadze}}, \bibfnamefont {G.}},
  and\ \bibinfo {author} {\bibnamefont {{Porrati}}, \bibfnamefont {M.}},\
  }\href {https://doi.org/10.1016/S0370-2693(00)00669-9} {\bibfield  {journal}
  {\bibinfo  {journal} {Physics Letters B}\ }\textbf {\bibinfo {volume}
  {485}},\ \bibinfo {pages} {208} (\bibinfo {year} {2000})},\ \Eprint
  {https://arxiv.org/abs/hep-th/0005016} {arXiv:hep-th/0005016 [hep-th]}
  \BibitemShut {NoStop}%
\bibitem [{\citenamefont {Einstein}(1917)}]{einstein1917kosmologische}%
  \BibitemOpen
  \bibfield  {author} {\bibinfo {author} {\bibnamefont {Einstein},
  \bibfnamefont {A.}},\ }\href@noop {} {\bibfield  {journal} {\bibinfo
  {journal} {SPA der Wissenschaften}\ }\textbf {\bibinfo {volume} {142}}
  (\bibinfo {year} {1917})}\BibitemShut {NoStop}%
\bibitem [{\citenamefont {{Eisenstein}}\ \emph {et~al.}(2005)\citenamefont
  {{Eisenstein}}, \citenamefont {{Zehavi}}, \citenamefont {{Hogg}},\ and\
  \citenamefont {othres}}]{2005ApJ...633..560E}%
  \BibitemOpen
  \bibfield  {author} {\bibinfo {author} {\bibnamefont {{Eisenstein}},
  \bibfnamefont {D.~J.}}, \bibinfo {author} {\bibnamefont {{Zehavi}},
  \bibfnamefont {I.}}, \bibinfo {author} {\bibnamefont {{Hogg}}, \bibfnamefont
  {D.~W.}}, and\ \bibinfo {author} {\bibnamefont {othres},},\ }\href
  {https://doi.org/10.1086/466512} {\bibfield  {journal} {\bibinfo  {journal}
  {\apj}\ }\textbf {\bibinfo {volume} {633}},\ \bibinfo {pages} {560} (\bibinfo
  {year} {2005})},\ \Eprint {https://arxiv.org/abs/astro-ph/0501171}
  {arXiv:astro-ph/0501171 [astro-ph]} \BibitemShut {NoStop}%
\bibitem [{\citenamefont {Englert}\ and\ \citenamefont
  {Brout}(1964)}]{PhysRevLett.13.321}%
  \BibitemOpen
  \bibfield  {author} {\bibinfo {author} {\bibnamefont {Englert}, \bibfnamefont
  {F.}}and\ \bibinfo {author} {\bibnamefont {Brout}, \bibfnamefont {R.}},\
  }\href {https://doi.org/10.1103/PhysRevLett.13.321} {\bibfield  {journal}
  {\bibinfo  {journal} {Phys. Rev. Lett.}\ }\textbf {\bibinfo {volume} {13}},\
  \bibinfo {pages} {321} (\bibinfo {year} {1964})}\BibitemShut {NoStop}%
\bibitem [{\citenamefont {Ezquiaga}\ and\ \citenamefont
  {Zumalac{\'a}rregui}(2017)}]{ezquiaga2017dark}%
  \BibitemOpen
  \bibfield  {author} {\bibinfo {author} {\bibnamefont {Ezquiaga},
  \bibfnamefont {J.~M.}}and\ \bibinfo {author} {\bibnamefont
  {Zumalac{\'a}rregui}, \bibfnamefont {M.}},\ }\href@noop {} {\bibfield
  {journal} {\bibinfo  {journal} {Physical review letters}\ }\textbf {\bibinfo
  {volume} {119}},\ \bibinfo {pages} {251304} (\bibinfo {year}
  {2017})}\BibitemShut {NoStop}%
\bibitem [{\citenamefont {{Ezquiaga}}\ and\ \citenamefont
  {{Zumalac{\'a}rregui}}(2018)}]{2018FrASS...5...44E}%
  \BibitemOpen
  \bibfield  {author} {\bibinfo {author} {\bibnamefont {{Ezquiaga}},
  \bibfnamefont {J.~M.}}and\ \bibinfo {author} {\bibnamefont
  {{Zumalac{\'a}rregui}}, \bibfnamefont {M.}},\ }\href
  {https://doi.org/10.3389/fspas.2018.00044} {\bibfield  {journal} {\bibinfo
  {journal} {Frontiers in Astronomy and Space Sciences}\ }\textbf {\bibinfo
  {volume} {5}},\ \bibinfo {eid} {44} (\bibinfo {year} {2018})},\ \Eprint
  {https://arxiv.org/abs/1807.09241} {arXiv:1807.09241 [astro-ph.CO]}
  \BibitemShut {NoStop}%
\bibitem [{\citenamefont {Friedman}(1922)}]{Friedman}%
  \BibitemOpen
  \bibfield  {author} {\bibinfo {author} {\bibnamefont {Friedman},
  \bibfnamefont {A.}},\ }\href@noop {} {\bibfield  {journal} {\bibinfo
  {journal} {Zeitschrift f{\"u}r Physik}\ }\textbf {\bibinfo {volume} {10}},\
  \bibinfo {pages} {377} (\bibinfo {year} {1922})}\BibitemShut {NoStop}%
\bibitem [{\citenamefont {{Gleyzes}}\ \emph {et~al.}(2015)\citenamefont
  {{Gleyzes}}, \citenamefont {{Langlois}}, \citenamefont {{Piazza}},\ and\
  \citenamefont {{Vernizzi}}}]{2015JCAP...02..018G}%
  \BibitemOpen
  \bibfield  {author} {\bibinfo {author} {\bibnamefont {{Gleyzes}},
  \bibfnamefont {J.}}, \bibinfo {author} {\bibnamefont {{Langlois}},
  \bibfnamefont {D.}}, \bibinfo {author} {\bibnamefont {{Piazza}},
  \bibfnamefont {F.}}, and\ \bibinfo {author} {\bibnamefont {{Vernizzi}},
  \bibfnamefont {F.}},\ }\href {https://doi.org/10.1088/1475-7516/2015/02/018}
  {\bibfield  {journal} {\bibinfo  {journal} {\jcap}\ }\textbf {\bibinfo
  {volume} {2015}},\ \bibinfo {eid} {018} (\bibinfo {year} {2015})},\ \Eprint
  {https://arxiv.org/abs/1408.1952} {arXiv:1408.1952 [astro-ph.CO]}
  \BibitemShut {NoStop}%
\bibitem [{\citenamefont {Higgs}(1964{\natexlab{a}})}]{PhysRevLett.13.508}%
  \BibitemOpen
  \bibfield  {author} {\bibinfo {author} {\bibnamefont {Higgs}, \bibfnamefont
  {P.~W.}},\ }\href {https://doi.org/10.1103/PhysRevLett.13.508} {\bibfield
  {journal} {\bibinfo  {journal} {Phys. Rev. Lett.}\ }\textbf {\bibinfo
  {volume} {13}},\ \bibinfo {pages} {508} (\bibinfo {year}
  {1964}{\natexlab{a}})}\BibitemShut {NoStop}%
\bibitem [{\citenamefont {Higgs}(1964{\natexlab{b}})}]{Higgs:1964ia}%
  \BibitemOpen
  \bibfield  {author} {\bibinfo {author} {\bibnamefont {Higgs}, \bibfnamefont
  {P.~W.}},\ }\href {https://doi.org/10.1016/0031-9163(64)91136-9} {\bibfield
  {journal} {\bibinfo  {journal} {Phys. Lett.}\ }\textbf {\bibinfo {volume}
  {12}},\ \bibinfo {pages} {132} (\bibinfo {year}
  {1964}{\natexlab{b}})}\BibitemShut {NoStop}%
\bibitem [{\citenamefont {Horndeski}(1974)}]{horndeski1974second}%
  \BibitemOpen
  \bibfield  {author} {\bibinfo {author} {\bibnamefont {Horndeski},
  \bibfnamefont {G.~W.}},\ }\href@noop {} {\bibfield  {journal} {\bibinfo
  {journal} {International Journal of Theoretical Physics}\ }\textbf {\bibinfo
  {volume} {10}},\ \bibinfo {pages} {363} (\bibinfo {year} {1974})},\ \bibinfo
  {note}
  {\url{https://link.springer.com/article/10.1007/BF01807638}}\BibitemShut
  {NoStop}%
\bibitem [{\citenamefont {{Ho{\v{r}}ava}}(2009)}]{2009PhRvD..79h4008H}%
  \BibitemOpen
  \bibfield  {author} {\bibinfo {author} {\bibnamefont {{Ho{\v{r}}ava}},
  \bibfnamefont {P.}},\ }\href {https://doi.org/10.1103/PhysRevD.79.084008}
  {\bibfield  {journal} {\bibinfo  {journal} {\prd}\ }\textbf {\bibinfo
  {volume} {79}},\ \bibinfo {eid} {084008} (\bibinfo {year} {2009})},\ \Eprint
  {https://arxiv.org/abs/0901.3775} {arXiv:0901.3775 [hep-th]} \BibitemShut
  {NoStop}%
\bibitem [{\citenamefont {{Hu}}\ and\ \citenamefont
  {{Sawicki}}(2007)}]{2007PhRvD..76f4004H}%
  \BibitemOpen
  \bibfield  {author} {\bibinfo {author} {\bibnamefont {{Hu}}, \bibfnamefont
  {W.}}and\ \bibinfo {author} {\bibnamefont {{Sawicki}}, \bibfnamefont {I.}},\
  }\href {https://doi.org/10.1103/PhysRevD.76.064004} {\bibfield  {journal}
  {\bibinfo  {journal} {\prd}\ }\textbf {\bibinfo {volume} {76}},\ \bibinfo
  {eid} {064004} (\bibinfo {year} {2007})},\ \Eprint
  {https://arxiv.org/abs/0705.1158} {arXiv:0705.1158 [astro-ph]} \BibitemShut
  {NoStop}%
\bibitem [{\citenamefont {Hunter}(2007)}]{Hunter:2007}%
  \BibitemOpen
  \bibfield  {author} {\bibinfo {author} {\bibnamefont {Hunter}, \bibfnamefont
  {J.~D.}},\ }\href {https://doi.org/10.1109/MCSE.2007.55} {\bibfield
  {journal} {\bibinfo  {journal} {Computing in Science \& Engineering}\
  }\textbf {\bibinfo {volume} {9}},\ \bibinfo {pages} {90} (\bibinfo {year}
  {2007})}\BibitemShut {NoStop}%
\bibitem [{\citenamefont {Kobayashi}, \citenamefont {Yamaguchi},\ and\
  \citenamefont {Yokoyama}(2011)}]{kobayashi2011generalized}%
  \BibitemOpen
  \bibfield  {author} {\bibinfo {author} {\bibnamefont {Kobayashi},
  \bibfnamefont {T.}}, \bibinfo {author} {\bibnamefont {Yamaguchi},
  \bibfnamefont {M.}}, and\ \bibinfo {author} {\bibnamefont {Yokoyama},
  \bibfnamefont {J.}},\ }\href@noop {} {\bibfield  {journal} {\bibinfo
  {journal} {Progress of Theoretical Physics}\ }\textbf {\bibinfo {volume}
  {126}},\ \bibinfo {pages} {511} (\bibinfo {year} {2011})}\BibitemShut
  {NoStop}%
\bibitem [{\citenamefont {{Koutsoumbas}}\ \emph {et~al.}(2018)\citenamefont
  {{Koutsoumbas}}, \citenamefont {{Ntrekis}}, \citenamefont
  {{Papantonopoulos}},\ and\ \citenamefont
  {{Saridakis}}}]{2018JCAP...02..003K}%
  \BibitemOpen
  \bibfield  {author} {\bibinfo {author} {\bibnamefont {{Koutsoumbas}},
  \bibfnamefont {G.}}, \bibinfo {author} {\bibnamefont {{Ntrekis}},
  \bibfnamefont {K.}}, \bibinfo {author} {\bibnamefont {{Papantonopoulos}},
  \bibfnamefont {E.}}, and\ \bibinfo {author} {\bibnamefont {{Saridakis}},
  \bibfnamefont {E.~N.}},\ }\href
  {https://doi.org/10.1088/1475-7516/2018/02/003} {\bibfield  {journal}
  {\bibinfo  {journal} {\jcap}\ }\textbf {\bibinfo {volume} {2018}},\ \bibinfo
  {eid} {003} (\bibinfo {year} {2018})},\ \Eprint
  {https://arxiv.org/abs/1704.08640} {arXiv:1704.08640 [gr-qc]} \BibitemShut
  {NoStop}%
\bibitem [{\citenamefont {{Leclercq}}(2015)}]{2015arXiv151204985L}%
  \BibitemOpen
  \bibfield  {author} {\bibinfo {author} {\bibnamefont {{Leclercq}},
  \bibfnamefont {F.}},\ }\href@noop {} {\bibfield  {journal} {\bibinfo
  {journal} {arXiv e-prints}\ ,\ \bibinfo {eid} {arXiv:1512.04985}} (\bibinfo
  {year} {2015})},\ \Eprint {https://arxiv.org/abs/1512.04985}
  {arXiv:1512.04985 [astro-ph.CO]} \BibitemShut {NoStop}%
\bibitem [{\citenamefont {Lombriser}\ and\ \citenamefont
  {Taylor}(2016)}]{lombriser2016breaking}%
  \BibitemOpen
  \bibfield  {author} {\bibinfo {author} {\bibnamefont {Lombriser},
  \bibfnamefont {L.}}and\ \bibinfo {author} {\bibnamefont {Taylor},
  \bibfnamefont {A.}},\ }\href@noop {} {\bibfield  {journal} {\bibinfo
  {journal} {Journal of Cosmology and Astroparticle Physics}\ }\textbf
  {\bibinfo {volume} {2016}},\ \bibinfo {pages} {031} (\bibinfo {year}
  {2016})}\BibitemShut {NoStop}%
\bibitem [{\citenamefont {{Nicolis}}\ and\ \citenamefont
  {{Rattazzi}}(2004)}]{2004JHEP...06..059N}%
  \BibitemOpen
  \bibfield  {author} {\bibinfo {author} {\bibnamefont {{Nicolis}},
  \bibfnamefont {A.}}and\ \bibinfo {author} {\bibnamefont {{Rattazzi}},
  \bibfnamefont {R.}},\ }\href {https://doi.org/10.1088/1126-6708/2004/06/059}
  {\bibfield  {journal} {\bibinfo  {journal} {Journal of High Energy Physics}\
  }\textbf {\bibinfo {volume} {2004}},\ \bibinfo {eid} {059} (\bibinfo {year}
  {2004})},\ \Eprint {https://arxiv.org/abs/hep-th/0404159}
  {arXiv:hep-th/0404159 [hep-th]} \BibitemShut {NoStop}%
\bibitem [{Note1()}]{Note1}%
  \BibitemOpen
  \bibinfo {note} {As the term suggests, it is the study of different
  gravitational theories within the framework of cosmology. On the other hand,
  we could also see the perspective in which we develop cosmology using
  gravitational theories. In that case we could use the terms \protect \textit
  {gravitological cosmology} or \protect \textit {gravitational cosmology}.
  These terms depend on what one is inspired from.}\BibitemShut {Stop}%
\bibitem [{Note2()}]{Note2}%
  \BibitemOpen
  \bibinfo {note} {Note that the name \protect \textit {functors of actions}
  can be also replaced by relations of actions or any other name that best
  describes these novel theories. We give this name as we currently understand
  this best describes these mathematical entities.}\BibitemShut {Stop}%
\bibitem [{Note3()}]{Note3}%
  \BibitemOpen
  \bibinfo {note} {\protect \url
  {https://github.com/lontelis/AofEFT}}\BibitemShut {NoStop}%
\bibitem [{\citenamefont {{Ntelis}}\ \emph {et~al.}(2018)\citenamefont
  {{Ntelis}}, \citenamefont {{Ealet}}, \citenamefont {{Escoffier}},
  \citenamefont {{Hamilton}}, \citenamefont {{Hawken}}, \citenamefont {{Le
  Goff}}, \citenamefont {{Rich}},\ and\ \citenamefont
  {{Tilquin}}}]{2018JCAP...12..014N}%
  \BibitemOpen
  \bibfield  {author} {\bibinfo {author} {\bibnamefont {{Ntelis}},
  \bibfnamefont {P.}}, \bibinfo {author} {\bibnamefont {{Ealet}}, \bibfnamefont
  {A.}}, \bibinfo {author} {\bibnamefont {{Escoffier}}, \bibfnamefont {S.}},
  \bibinfo {author} {\bibnamefont {{Hamilton}}, \bibfnamefont {J.-C.}},
  \bibinfo {author} {\bibnamefont {{Hawken}}, \bibfnamefont {A.~J.}}, \bibinfo
  {author} {\bibnamefont {{Le Goff}}, \bibfnamefont {J.-M.}}, \bibinfo {author}
  {\bibnamefont {{Rich}}, \bibfnamefont {J.}}, and\ \bibinfo {author}
  {\bibnamefont {{Tilquin}}, \bibfnamefont {A.}},\ }\href
  {https://doi.org/10.1088/1475-7516/2018/12/014} {\bibfield  {journal}
  {\bibinfo  {journal} {\jcap}\ }\textbf {\bibinfo {volume} {2018}},\ \bibinfo
  {eid} {014} (\bibinfo {year} {2018})},\ \Eprint
  {https://arxiv.org/abs/1810.09362} {arXiv:1810.09362 [astro-ph.CO]}
  \BibitemShut {NoStop}%
\bibitem [{\citenamefont {{Ntelis}}\ \emph {et~al.}(2017)\citenamefont
  {{Ntelis}}, \citenamefont {{Hamilton}}, \citenamefont {{Le Goff}} \emph
  {et~al.}}]{2017JCAP...06..019N}%
  \BibitemOpen
  \bibfield  {author} {\bibinfo {author} {\bibnamefont {{Ntelis}},
  \bibfnamefont {P.}}, \bibinfo {author} {\bibnamefont {{Hamilton}},
  \bibfnamefont {J.-C.}}, \bibinfo {author} {\bibnamefont {{Le Goff}},
  \bibfnamefont {J.-M.}},  \emph {et~al.},\ }\href
  {https://doi.org/10.1088/1475-7516/2017/06/019} {\bibfield  {journal}
  {\bibinfo  {journal} {\jcap}\ }\textbf {\bibinfo {volume} {2017}},\ \bibinfo
  {eid} {019} (\bibinfo {year} {2017})},\ \Eprint
  {https://arxiv.org/abs/1702.02159} {arXiv:1702.02159 [astro-ph.CO]}
  \BibitemShut {NoStop}%
\bibitem [{\citenamefont {{Overduin}}\ and\ \citenamefont
  {{Wesson}}(1997)}]{1997PhR...283..303O}%
  \BibitemOpen
  \bibfield  {author} {\bibinfo {author} {\bibnamefont {{Overduin}},
  \bibfnamefont {J.~M.}}and\ \bibinfo {author} {\bibnamefont {{Wesson}},
  \bibfnamefont {P.~S.}},\ }\href
  {https://doi.org/10.1016/S0370-1573(96)00046-4} {\bibfield  {journal}
  {\bibinfo  {journal} {\physrep}\ }\textbf {\bibinfo {volume} {283}},\
  \bibinfo {pages} {303} (\bibinfo {year} {1997})},\ \Eprint
  {https://arxiv.org/abs/gr-qc/9805018} {arXiv:gr-qc/9805018 [gr-qc]}
  \BibitemShut {NoStop}%
\bibitem [{\citenamefont {{Pajer}}\ and\ \citenamefont
  {{Zaldarriaga}}(2013)}]{2013JCAP...08..037P}%
  \BibitemOpen
  \bibfield  {author} {\bibinfo {author} {\bibnamefont {{Pajer}}, \bibfnamefont
  {E.}}and\ \bibinfo {author} {\bibnamefont {{Zaldarriaga}}, \bibfnamefont
  {M.}},\ }\href {https://doi.org/10.1088/1475-7516/2013/08/037} {\bibfield
  {journal} {\bibinfo  {journal} {\jcap}\ }\textbf {\bibinfo {volume} {2013}},\
  \bibinfo {eid} {037} (\bibinfo {year} {2013})},\ \Eprint
  {https://arxiv.org/abs/1301.7182} {arXiv:1301.7182 [astro-ph.CO]}
  \BibitemShut {NoStop}%
\bibitem [{\citenamefont {Perenon}, \citenamefont {Marinoni},\ and\
  \citenamefont {Piazza}(2017)}]{PerenonMarinoniPiazza}%
  \BibitemOpen
  \bibfield  {author} {\bibinfo {author} {\bibnamefont {Perenon}, \bibfnamefont
  {L.}}, \bibinfo {author} {\bibnamefont {Marinoni}, \bibfnamefont {C.}}, and\
  \bibinfo {author} {\bibnamefont {Piazza}, \bibfnamefont {F.}},\ }\href
  {http://stacks.iop.org/1475-7516/2017/i=01/a=035} {\bibfield  {journal}
  {\bibinfo  {journal} {Journal of Cosmology and Astroparticle Physics}\
  }\textbf {\bibinfo {volume} {2017}},\ \bibinfo {pages} {035} (\bibinfo {year}
  {2017})}\BibitemShut {NoStop}%
\bibitem [{\citenamefont {{Perez}}\ and\ \citenamefont
  {{Granger}}(2007)}]{4160251}%
  \BibitemOpen
  \bibfield  {author} {\bibinfo {author} {\bibnamefont {{Perez}}, \bibfnamefont
  {F.}}and\ \bibinfo {author} {\bibnamefont {{Granger}}, \bibfnamefont
  {B.~E.}},\ }\href@noop {} {\bibfield  {journal} {\bibinfo  {journal}
  {Computing in Science Engineering}\ }\textbf {\bibinfo {volume} {9}},\
  \bibinfo {pages} {21} (\bibinfo {year} {2007})}\BibitemShut {NoStop}%
\bibitem [{\citenamefont {Peskin}\ and\ \citenamefont
  {Schroeder}(1995)}]{Peskin:1995ev}%
  \BibitemOpen
  \bibfield  {author} {\bibinfo {author} {\bibnamefont {Peskin}, \bibfnamefont
  {M.~E.}}and\ \bibinfo {author} {\bibnamefont {Schroeder}, \bibfnamefont
  {D.~V.}},\ }\href@noop {} {\emph {\bibinfo {title} {{An Introduction to
  quantum field theory}}}}\ (\bibinfo  {publisher} {Addison-Wesley},\ \bibinfo
  {address} {Reading, USA},\ \bibinfo {year} {1995})\BibitemShut {NoStop}%
\bibitem [{\citenamefont {{Piazza}}\ and\ \citenamefont
  {{Vernizzi}}(2013)}]{2013CQGra..30u4007P}%
  \BibitemOpen
  \bibfield  {author} {\bibinfo {author} {\bibnamefont {{Piazza}},
  \bibfnamefont {F.}}and\ \bibinfo {author} {\bibnamefont {{Vernizzi}},
  \bibfnamefont {F.}},\ }\href {https://doi.org/10.1088/0264-9381/30/21/214007}
  {\bibfield  {journal} {\bibinfo  {journal} {Classical and Quantum Gravity}\
  }\textbf {\bibinfo {volume} {30}},\ \bibinfo {eid} {214007} (\bibinfo {year}
  {2013})},\ \Eprint {https://arxiv.org/abs/1307.4350} {arXiv:1307.4350
  [hep-th]} \BibitemShut {NoStop}%
\bibitem [{\citenamefont {{Planck Collaboration}}\ \emph
  {et~al.}(2020)\citenamefont {{Planck Collaboration}}, \citenamefont
  {{Aghanim}}, \citenamefont {{Akrami}}, \citenamefont {{Ashdown}} \emph
  {et~al.}}]{2020A&A...641A...6P}%
  \BibitemOpen
  \bibfield  {author} {\bibinfo {author} {\bibnamefont {{Planck
  Collaboration}},}, \bibinfo {author} {\bibnamefont {{Aghanim}}, \bibfnamefont
  {N.}}, \bibinfo {author} {\bibnamefont {{Akrami}}, \bibfnamefont {Y.}},
  \bibinfo {author} {\bibnamefont {{Ashdown}}, \bibfnamefont {M.}},  \emph
  {et~al.},\ }\href {https://doi.org/10.1051/0004-6361/201833910} {\bibfield
  {journal} {\bibinfo  {journal} {\aap}\ }\textbf {\bibinfo {volume} {641}},\
  \bibinfo {eid} {A6} (\bibinfo {year} {2020})},\ \Eprint
  {https://arxiv.org/abs/1807.06209} {arXiv:1807.06209 [astro-ph.CO]}
  \BibitemShut {NoStop}%
\bibitem [{\citenamefont {Polyakov}(1981)}]{polyakov1981quantum}%
  \BibitemOpen
  \bibfield  {author} {\bibinfo {author} {\bibnamefont {Polyakov},
  \bibfnamefont {A.~M.}},\ }\href@noop {} {\bibfield  {journal} {\bibinfo
  {journal} {Physics Letters B}\ }\textbf {\bibinfo {volume} {103}},\ \bibinfo
  {pages} {207} (\bibinfo {year} {1981})}\BibitemShut {NoStop}%
\bibitem [{\citenamefont {{Porto}}(2016)}]{2016PhR...633....1P}%
  \BibitemOpen
  \bibfield  {author} {\bibinfo {author} {\bibnamefont {{Porto}}, \bibfnamefont
  {R.~A.}},\ }\href {https://doi.org/10.1016/j.physrep.2016.04.003} {\bibfield
  {journal} {\bibinfo  {journal} {\physrep}\ }\textbf {\bibinfo {volume}
  {633}},\ \bibinfo {pages} {1} (\bibinfo {year} {2016})},\ \Eprint
  {https://arxiv.org/abs/1601.04914} {arXiv:1601.04914 [hep-th]} \BibitemShut
  {NoStop}%
\bibitem [{\citenamefont {{Randall}}\ and\ \citenamefont
  {{Sundrum}}(1999)}]{1999PhRvL..83.3370R}%
  \BibitemOpen
  \bibfield  {author} {\bibinfo {author} {\bibnamefont {{Randall}},
  \bibfnamefont {L.}}and\ \bibinfo {author} {\bibnamefont {{Sundrum}},
  \bibfnamefont {R.}},\ }\href {https://doi.org/10.1103/PhysRevLett.83.3370}
  {\bibfield  {journal} {\bibinfo  {journal} {\prl}\ }\textbf {\bibinfo
  {volume} {83}},\ \bibinfo {pages} {3370} (\bibinfo {year} {1999})},\ \Eprint
  {https://arxiv.org/abs/hep-ph/9905221} {arXiv:hep-ph/9905221 [hep-ph]}
  \BibitemShut {NoStop}%
\bibitem [{\citenamefont {{Seiberg}}\ and\ \citenamefont
  {{Witten}}(1999)}]{1999JHEP...09..032S}%
  \BibitemOpen
  \bibfield  {author} {\bibinfo {author} {\bibnamefont {{Seiberg}},
  \bibfnamefont {N.}}and\ \bibinfo {author} {\bibnamefont {{Witten}},
  \bibfnamefont {E.}},\ }\href {https://doi.org/10.1088/1126-6708/1999/09/032}
  {\bibfield  {journal} {\bibinfo  {journal} {Journal of High Energy Physics}\
  }\textbf {\bibinfo {volume} {1999}},\ \bibinfo {eid} {032} (\bibinfo {year}
  {1999})},\ \Eprint {https://arxiv.org/abs/hep-th/9908142}
  {arXiv:hep-th/9908142 [hep-th]} \BibitemShut {NoStop}%
\bibitem [{\citenamefont {{Sotiriou}}\ and\ \citenamefont
  {{Faraoni}}(2010)}]{2010RvMP...82..451S}%
  \BibitemOpen
  \bibfield  {author} {\bibinfo {author} {\bibnamefont {{Sotiriou}},
  \bibfnamefont {T.~P.}}and\ \bibinfo {author} {\bibnamefont {{Faraoni}},
  \bibfnamefont {V.}},\ }\href {https://doi.org/10.1103/RevModPhys.82.451}
  {\bibfield  {journal} {\bibinfo  {journal} {Reviews of Modern Physics}\
  }\textbf {\bibinfo {volume} {82}},\ \bibinfo {pages} {451} (\bibinfo {year}
  {2010})},\ \Eprint {https://arxiv.org/abs/0805.1726} {arXiv:0805.1726
  [gr-qc]} \BibitemShut {NoStop}%
\bibitem [{\citenamefont {Starobinsky}(1980)}]{STAROBINSKY198099}%
  \BibitemOpen
  \bibfield  {author} {\bibinfo {author} {\bibnamefont {Starobinsky},
  \bibfnamefont {A.}},\ }\href
  {https://doi.org/https://doi.org/10.1016/0370-2693(80)90670-X} {\bibfield
  {journal} {\bibinfo  {journal} {Physics Letters B}\ }\textbf {\bibinfo
  {volume} {91}},\ \bibinfo {pages} {99 } (\bibinfo {year} {1980})}\BibitemShut
  {NoStop}%
\bibitem [{\citenamefont {{Virtanen}}\ \emph {et~al.}(2019)\citenamefont
  {{Virtanen}}, \citenamefont {{Gommers}}, \citenamefont {{Oliphant}},\ and\
  \citenamefont {{Contributors}}}]{2019arXiv190710121V}%
  \BibitemOpen
  \bibfield  {author} {\bibinfo {author} {\bibnamefont {{Virtanen}},
  \bibfnamefont {P.}}, \bibinfo {author} {\bibnamefont {{Gommers}},
  \bibfnamefont {R.}}, \bibinfo {author} {\bibnamefont {{Oliphant}},
  \bibfnamefont {T.~E.}}, and\ \bibinfo {author} {\bibnamefont
  {{Contributors}}, \bibfnamefont {S.~.~.}},\ }\href@noop {} {\bibfield
  {journal} {\bibinfo  {journal} {arXiv e-prints}\ ,\ \bibinfo {eid}
  {arXiv:1907.10121}} (\bibinfo {year} {2019})},\ \Eprint
  {https://arxiv.org/abs/1907.10121} {arXiv:1907.10121 [cs.MS]} \BibitemShut
  {NoStop}%
\bibitem [{\citenamefont {Walt}, \citenamefont {Colbert},\ and\ \citenamefont
  {Varoquaux}(2011)}]{Walt:2011:NAS:1957373.1957466}%
  \BibitemOpen
  \bibfield  {author} {\bibinfo {author} {\bibnamefont {Walt}, \bibfnamefont
  {S.~v.~d.}}, \bibinfo {author} {\bibnamefont {Colbert}, \bibfnamefont
  {S.~C.}}, and\ \bibinfo {author} {\bibnamefont {Varoquaux}, \bibfnamefont
  {G.}},\ }\href {https://doi.org/10.1109/MCSE.2011.37} {\bibfield  {journal}
  {\bibinfo  {journal} {Computing in Science and Engg.}\ }\textbf {\bibinfo
  {volume} {13}},\ \bibinfo {pages} {22} (\bibinfo {year} {2011})}\BibitemShut
  {NoStop}%
\bibitem [{\citenamefont {{Wikipedia
  contributors}}(2018)}]{wiki:CovariantDerivative}%
  \BibitemOpen
  \bibfield  {author} {\bibinfo {author} {\bibnamefont {{Wikipedia
  contributors}},},\ }\href
  {https://en.wikipedia.org/w/index.php?title=Covariant_derivative&oldid=842130744}
  {\enquote {\bibinfo {title} {Covariant derivative --- {Wikipedia}{,} the free
  encyclopedia},}\ } (\bibinfo {year} {2018}),\ \bibinfo {note} {[Online;
  accessed 16-August-2018]}\BibitemShut {NoStop}%
\bibitem [{\citenamefont {Wilks}(1938)}]{wilks}%
  \BibitemOpen
  \bibfield  {author} {\bibinfo {author} {\bibnamefont {Wilks}, \bibfnamefont
  {S.~S.}},\ }\href {https://doi.org/10.1214/aoms/1177732360} {\bibfield
  {journal} {\bibinfo  {journal} {The Annals of Mathematical Statistics}\
  }\textbf {\bibinfo {volume} {9}},\ \bibinfo {pages} {60 } (\bibinfo {year}
  {1938})}\BibitemShut {NoStop}%
\bibitem [{\citenamefont {{Witten}}(1998)}]{1998AdTMP...2..253W}%
  \BibitemOpen
  \bibfield  {author} {\bibinfo {author} {\bibnamefont {{Witten}},
  \bibfnamefont {E.}},\ }\href@noop {} {\bibfield  {journal} {\bibinfo
  {journal} {Advances in Theoretical and Mathematical Physics}\ }\textbf
  {\bibinfo {volume} {2}},\ \bibinfo {pages} {253} (\bibinfo {year} {1998})},\
  \Eprint {https://arxiv.org/abs/hep-th/9802150} {arXiv:hep-th/9802150
  [hep-th]} \BibitemShut {NoStop}%
\end{thebibliography}%
